\documentclass[a4paper,11pt]{article}
\pdfoutput=1
\usepackage{jcappub}
\usepackage{bm}
\usepackage{soul}
\usepackage{latexsym}
\usepackage{dcolumn}
\usepackage{amsfonts,amssymb,amsmath}
\usepackage{graphicx,epsfig}
\usepackage{psfrag}
\usepackage{braket}
\usepackage{subfigure}
\usepackage{rotating}
\usepackage{hyperref}
\usepackage{tikz}

\hypersetup{
	unicode=false,          
	pdftoolbar=true,        
	pdfmenubar=true,        
	pdffitwindow=false,     
	pdfstartview={FitH},    
	pdftitle={My title},    
	pdfauthor={Author},     
	pdfsubject={Subject},   
	pdfcreator={Creator},   
	pdfproducer={Producer}, 
	pdfkeywords={keyword1} {key2} {key3}, 
	pdfnewwindow=true,      
	colorlinks=true,       
	linkcolor=red,          
	citecolor=cyan,        
	filecolor=magenta,      
	urlcolor=green,           
	linktocpage=true
}

\makeatletter
\gdef\@fpheader{}
\makeatother

\begin{document}
\title{Back to the features: assessing the discriminating power of future CMB missions on inflationary models }

\author[1,2,3]{Matteo~Braglia,}
\author[4]{Xingang Chen,}
\author[5,6,3]{Dhiraj Kumar Hazra,}
\author[1]{Lucas Pinol}

\affiliation[1]{Instituto de Fisica Teorica, Universidad Autonoma de Madrid, Madrid, 28049, Spain}

\affiliation[2]{Center for Cosmology and Particle Physics, New York University, 726 Broadway, New York, NY 10003, USA}
\affiliation[3]{INAF/OAS Bologna, via Gobetti 101, I-40129 Bologna, Italy}

\affiliation[4]{Institute for Theory and Computation, Harvard-Smithsonian Center for Astrophysics, 60
	Garden Street, Cambridge, MA 02138, USA}
\affiliation[5]{The  Institute  of  Mathematical  Sciences,  HBNI,  CIT  Campus, Chennai  600113,  India}
\affiliation[6]{Homi Bhabha National Institute, Training School Complex, Anushakti Nagar, Mumbai 400085, India}

\emailAdd{mb9289@nyu.edu,
	xingang.chen@cfa.harvard.edu,
	dhiraj@imsc.res.in,
	lucas.pinol@ift.csic.es}

\abstract
{Future Cosmic Microwave Background (CMB) experiments  will deliver extremely accurate measurements of the E-modes pattern of the CMB polarization field.
Given the sharpness of the E-modes transfer functions, such surveys make for a powerful detector of high-frequency signals from  primordial features that may be lurking in current data sets.
With a handful of toy models that increase the fit to the latest Planck data, but are of marginal statistical significance, we use a state-of-the-art forecast pipeline to illustrate the promising prospects to test primordial features in the next decade. 
Not only will future experiments allow us to detect such features in data, but they will also be able to discriminate between models and narrow down the physical mechanism originating them with high statistical significance.
On the other hand, if the anomalies in the currently measured CMB spectra are just statistical fluctuations,  all the current feature best fit candidates will be ruled out. Either way, our results show that primordial features are a clear target of forthcoming CMB surveys beyond the detection of tensor modes. }

\maketitle

		\newpage
		\setcounter{page}{1}
		
		\tableofcontents
		

		\section{Introduction}
		\label{Sec:Introduction}
		\setcounter{equation}{0}

		Precision cosmological observations set tight constraints on the shape of the primordial power spectrum (PPS) from inflation. The most stringent ones come from the latest release by the Planck satellite, which produced the most accurate maps to date of $T$ and $E$ modes of the Cosmic Microwave Background (CMB) anisotropies from largest to small scales~\cite{Planck:2018nkj}. Planck probes scales $k\in[10^{-3},\,10^{-1}]\, {\rm Mpc}^{-1}$, corresponding to physical wavelength being stretched above the Hubble radius, for example, $\sim60-55$ $e$-folds before the end of inflation models with very high energy scales.
		At those scales, the analysis of Planck data favors a parameterization of the power spectrum of the curvature perturbation $\zeta$ with only 2 parameters, $\mathcal{P}_0(k)\equiv \lvert\zeta(k)\rvert^2\,k^3/2\pi^2=A_s \left(k/k_*\right)^{n_s-1}$, and accurately constrains them to  $\ln(10^{10}\, A_s)=3.045 \pm 0.016$ and $n_s = 0.9649 \pm 0.0044$ at 68\% CL, using temperature and E-mode polarization data~\cite{Planck:2018jri}.
		Here $k_*$ is called the pivot scale, and is conventionally set to $0.05\, {\rm Mpc}^{-1}$.

\begin{figure}
	\begin{center}
		\includegraphics[width=\columnwidth]{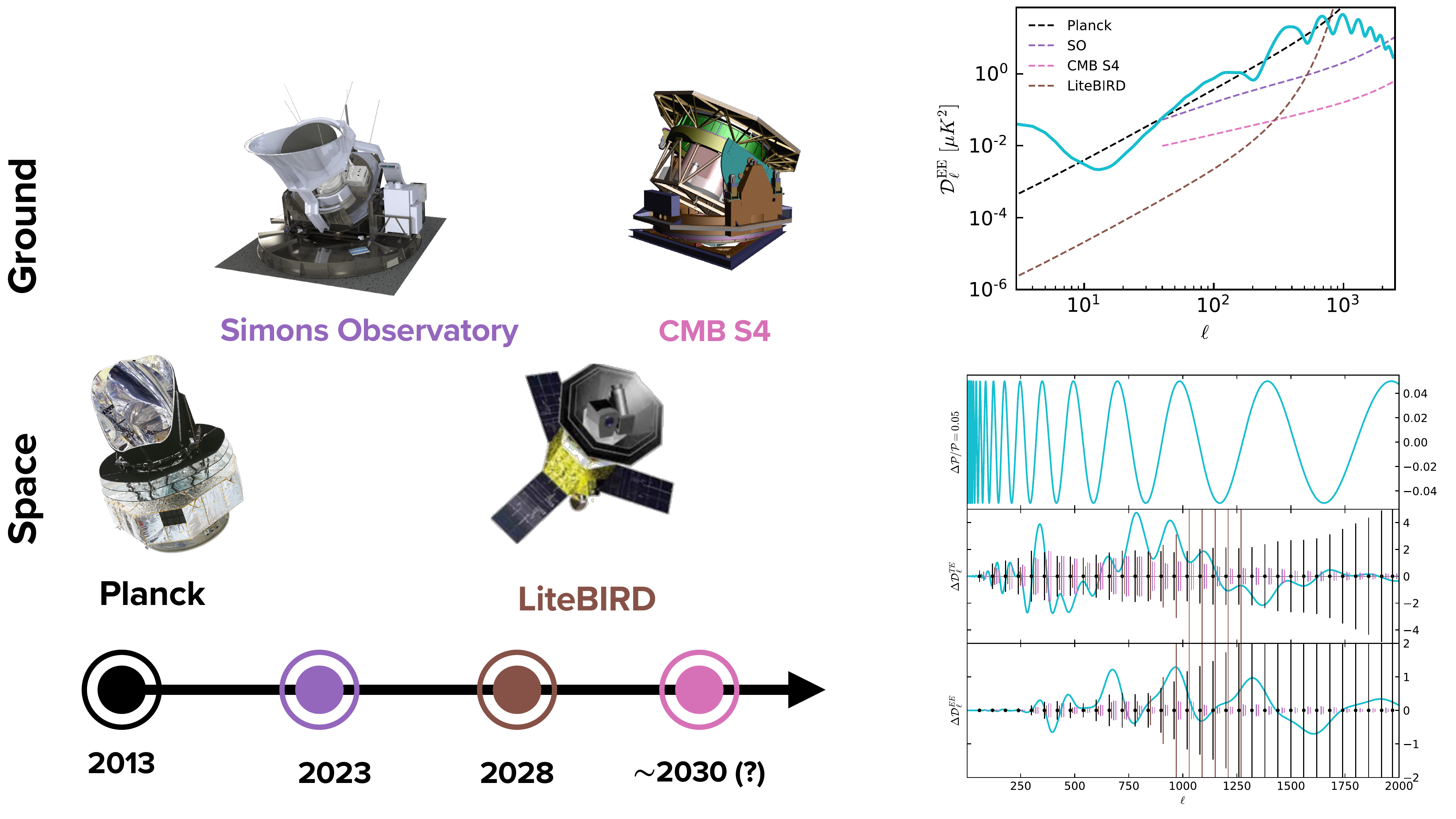}\end{center}
	\caption{\footnotesize\label{tab:noise}  Timeline for future CMB experiments (left panel), their E-modes noise curves (upper right panel) and associated error bars (lower right panel).
	In the lower right panel, we also show the signal (blue line) resulting from a resonant feature model with an amplitude $\Delta \mathcal{P}/\mathcal{P}=0.05$, typically of the order of magnitude of the features hinted at by Planck data.
	Note also that the error bars have been centered around zero for illustration purposes only, so they can be easily compared to the amplitude of the signal.
	}
\end{figure}

		Simplest models of inflation are said to be of the Single-Field Slow-Roll (SFSR) kind: a scalar field slowly rolls down a sufficiently flat potential, thereby producing a spectrum of primordial fluctuations close to scale invariance.
		The exact value of the deviation from scale invariance, encoded in the parameter $n_s-1$, varies from model to model.
		Impressively, current data is precise enough to discriminate models of SFSR inflation based on the favored value for the spectral index $n_s$.
		In fact, another key prediction of inflation is the production of a background of gravitational waves~\cite{Starobinsky:1979ty}, often parameterized by $r$ as the ratio of the amplitude of their spectrum to that of the scalar perturbations, for which non-observations of primordial B-modes from BICEP-Keck and Planck set the stringent upper limit $r_{0.05 {\rm Mpc}^{-1}}<0.036$ at 95\% CL~\cite{BICEP:2021xfz} (see also~\cite{Campeti:2022vom,Tristram:2021tvh,Paoletti:2022anb}).
		Joint constraints on $(n_s,r)$ favor models with concave potentials featuring a low $r$, such as the original proposal of $R^2$-inflation~\cite{Starobinsky:1980te}, its $\alpha$-attractor generalizations~\cite{Kallosh:2013yoa,Carrasco:2015uma}, or small-field inflation models.
		The hope for a detection of primordial gravitational waves have driven efforts towards building next-generation experiments with greatly improved sensitivity to polarization anisotropies of the CMB.
		In particular, primordial $B$ modes will be probed at an unprecedented level of precision, with forecast sensitivities on $r$ down to the level of $r\lesssim \mathcal{O}(10^{-3})$~\cite{Kallosh:2019eeu}.
		Such experiments include the Simons Observatory~\cite{SimonsObservatory:2018koc}, LiteBIRD~\cite{LiteBIRD:2020khw} and  CMB-S4~\cite{CMB-S4:2016ple}. Their respective (tentative) timelines are displayed in Fig.~\ref{tab:noise}.
		While several motivated models of inflation predict $r$ to be in the ballpark of future sensitivities, our ignorance about
		the physics at the very high energies does not allow us to make definitive predictions.
		In fact, there are small-field realizations of Slow-Roll (SR) that predict $r$ to be even orders of magnitude smaller than this.
		It is therefore of utmost importance to consider observables beyond $(n_s,r)$, in order to extract as much information as possible from the high-quality data that future experiments will deliver in the forthcoming years.

		Primordial features provide an example of such observables~\cite{Chen:2010xka,Chluba:2015bqa,Slosar:2019gvt,Achucarro:2022qrl}. They are scale-dependent corrections induced by temporary or small deviations from the SR evolution of the inflationary background.
		Such deviations can be caused by features in the inflationary potential, in the internal field space or in the couplings between the inflaton and other fields active during inflation.
		Broadly, one can distinguish two families of features: sharp and resonant features~\cite{Chen:2008wn}, depending on whether the deviation from SR is nearly instantaneous or periodic.
		The resulting modulation of the (otherwise near-scale-invariant) power spectrum shows up as an oscillatory pattern with a characteristic running of the frequency, as well as a model-dependent envelope.
		Specifically, sharp features generate a correction of the form $\Delta \mathcal{P}_{\rm sharp}(k)\propto\mathcal{P}_0 \sin (k/k_0)$ where $k_0$ is the mode that crosses the horizon at the time of the sharp feature, while resonant features take the form $\Delta \mathcal{P}_{\rm res}(k)\propto\mathcal{P}_0 \sin (\omega \ln k + {\rm phase})$, where the running is linear in $\ln k$ space.
		Some models of inflation, such as the so-called Classical Primordial Standard Clocks (CPSC) predict a complicated superposition of the two kinds of features \cite{Chen:2011zf, Chen:2011tu, Chen:2012ja, Battefeld:2013xka, Gao:2013ota, Noumi:2013cfa, Saito:2012pd, Saito:2013aqa, Chen:2014joa, Chen:2014cwa, Huang:2016quc, Domenech:2018bnf, Wang:2020aqc, Braglia:2021ckn, Braglia:2021sun, Braglia:2021rej, Hamann:2021eyw, Bodas:2022zca, Chen:2022vzh}.\footnote{Recently, it has also been pointed out that an oscillatory feature in the inflaton potential, with a bumpy modulation may address some interesting residuals in CMB data~\cite{Antony:2021bgp,Hazra:2022rdl,Antony:2022ert}. In these models, in general, only one of the two runnings shows up clearly and the other acts as a modulation of the dominant running.}
		Measuring the parameters describing $\Delta \mathcal{P}$ would give access to the fundamental physics at the origin of the signal, necessarily beyond the simple SFSR paradigm.

        Interestingly, current data may already hint at primordial features.
        Temperature data indeed show residuals that do not agree with the standard cosmological model, with a statistical significance up to $3\sigma$, and that can be explained by primordial features.
        Such {\em anomalies} include a systematic suppression of power in the $TT$ angular spectra compared to the $\Lambda$CDM best-fit for large scales, $\ell\lesssim 30$, and an oscillatory-like structure of residuals at higher $\ell$, which is mostly significant around $\ell\sim750$.
        Many feature models have been compared to Planck data in the literature (see \cite{Slosar:2019gvt,Achucarro:2022qrl} for reviews).
        Although interesting best-fit candidates have been identified, none of the feature models is statistically favored over the SFSR paradigm, due to the introduction of penalizing extra parameters.
        
         Given the hints for them from the data, and the interesting physical information that would be extracted by their detection, the purpose of this paper is to consolidate the status of features as a target for future CMB experiments.
         Indeed, although those experiments were mainly designed for the detection of primordial $B$ modes, they will also achieve an unprecedented precision in the measurement of $E$ modes. 
         As shown in Fig.~\ref{tab:noise}, this will result in reducing the error bars on $EE$ power spectra (and $TE$ as well) to the level of cosmic variance, which will be much more sensitive to features than temperature spectra, due to their narrower transfer functions in $\ell$ space.
         Future CMB experiments  therefore offer new prospects for assessing the statistical significance of features by jointly constraining their correlated effects on $T$ and $E$ modes, as also noticed in Ref.~\cite{CORE:2016ymi}.
         
        Specifically,  the purpose of this paper is to address the two following questions. Will we be able to rule out the above mentioned feature models if they are truly only statistical fluctuations, and conversely detect them if they are not? Supposing that a feature is detected in a future experiment, how well can we distinguish between different feature models?
        In particular, the second question has never been addressed, while it would crucially tell us to which extent features can be used to constrain the physics of inflation.
        To meet our goals, we use an efficient pipeline for the analysis of primordial feature models, introduced in Ref.~\cite{Braglia:2021ckn}.
        We consider several feature models that improve the fit to current Planck data and we forecast the capability of future experiments to tell them apart.
        Rather than using templates for the feature signal, using BI-spectra and Non-Gaussianity Operator (BINGO)~\cite{Hazra:2012yn} and its extension to two-field models~\cite{Braglia:2020fms}, we numerically integrate the equations for linear perturbations and arrive at exact power spectra, in order to fully capture model-dependent details that may be overlooked using analytical approximations.
        This enables us to explore fine details of the signal, that such experiments will be sensitive to.
        In our analyses, we pay particular attention to the implications of a future detection on early Universe model building.
        Our results are very promising and generically prove that we will be able to assess the statistical significance of primordial feature models within the next decade of CMB observations.

        Our paper is structured as follows.
        We start in Sec.~\ref{sec:models} with a description of the feature models that we use in our analysis, and present their corresponding best-fit.
        In Sec.~\ref{sec:featureless}, we assume that hints for features constitute statistical fluctuations in the current data and that the true model of the universe is featureless; we forecast the corresponding constraints on feature models that we would get.
        In Sec.~\ref{sec:features}, we answer the opposite question, i.e. we assume for each of these models that it provides the correct description of the Universe, and present the constraints on its primordial power spectrum that we would get in that case.
        Sec.~\ref{sec:comparison1} is instead devoted to the question of whether we can distinguish amongst different feature models if a feature is detected in the data. For this, we assume CPSC as a fiducial model for the Universe and discuss how much we could tell apart this model from other feature models. Taking advantage of the properties of CPSC, we also discuss the prospects for ruling out alternative scenarios to inflation.
        Sec.~\ref{sec:comparison2} is dedicated to the particular case of feature models that predict very similar angular power spectra: we argue that primordial non-Gaussianities may be used as an additional source of information to disentangle feature models.
        In Sec.~\ref{sec:estimator}, we introduce a simple estimator that can be used to quickly quantify the difference between two feature signals and roughly assess the prospects for distinguishing them in light of our results without further data analyses.  We discuss the implications of our results in Sec.~\ref{sec:conclusions}.
        
        While our paper aims at being self-consistent, the discussion could be hard to follow if we described every step of our analysis in the main text. Therefore, to facilitate the reading, we collect all technicalities in Appendices.
        In App.~\ref{app:analysis}, we describe our forecast pipeline and methodology.
        In App.~\ref{app:analytical_results} we analytically estimate the feature signal for some models to motivate the choice of the effective parameters used in our analysis. In App.~\ref{app:cmb_estimates}, we give a rough analytical  estimate of the imprints of primordial features on T and E spectra to show their sensitivity to different runnings and frequencies of the signals. In our paper, we use $\ln$ and $\log$ interchangeably to denote natural logarithm and we explicitly write $\log_{10}$ to denote the logarithm in basis 10.

		\section{The models}
		
		\label{sec:models}

		\begin{figure*}
			\includegraphics[width=.495\columnwidth]{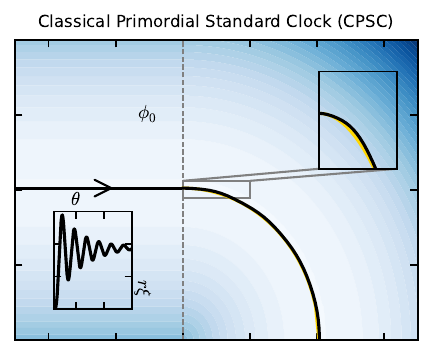}
			\includegraphics[width=.495\columnwidth]{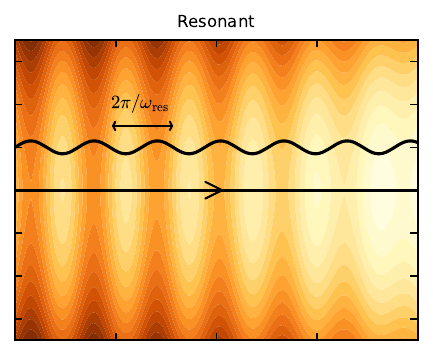}
			\includegraphics[width=.495\columnwidth]{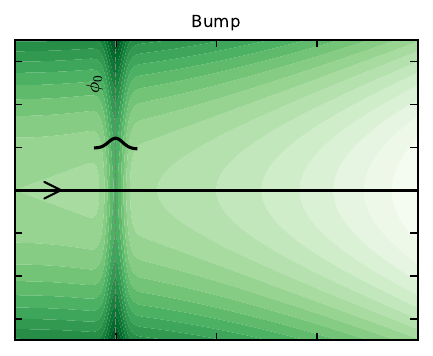}
			\includegraphics[width=.495\columnwidth]{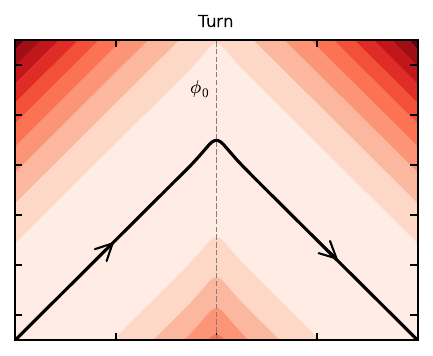}
			\caption{\footnotesize\label{fig:Trajectory} Inflationary trajectories considered in our work.
			[Top-Left, CPSC model] The inflationary trajectory eventually enters a curved path of valley that excites the oscillations of a massive field (see insert). [Top-Right, Resonant model] The trajectory is straight, but the inflaton potential features oscillatory modulations. [Bottom-Left, Bump/Dip model] The straight trajectory is shortly disturbed by a bump (or dip, not shown in the figure) in the inflaton potential. [Bottom-Right, Turn model] Two straight inflationary trajectories are connected by a sharp turn.
   These plots are for illustrative purposes only, indeed the features consistent with real data correspond to modifications in the trajectory that are not visible by eye, we have therefore voluntarily omitted quantitative labels.
   }
		\end{figure*}

		\begin{figure*}
			\includegraphics[width=.4\columnwidth]{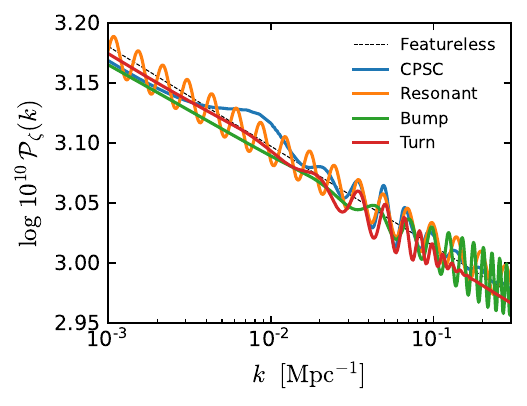}
			\includegraphics[width=.6\columnwidth]{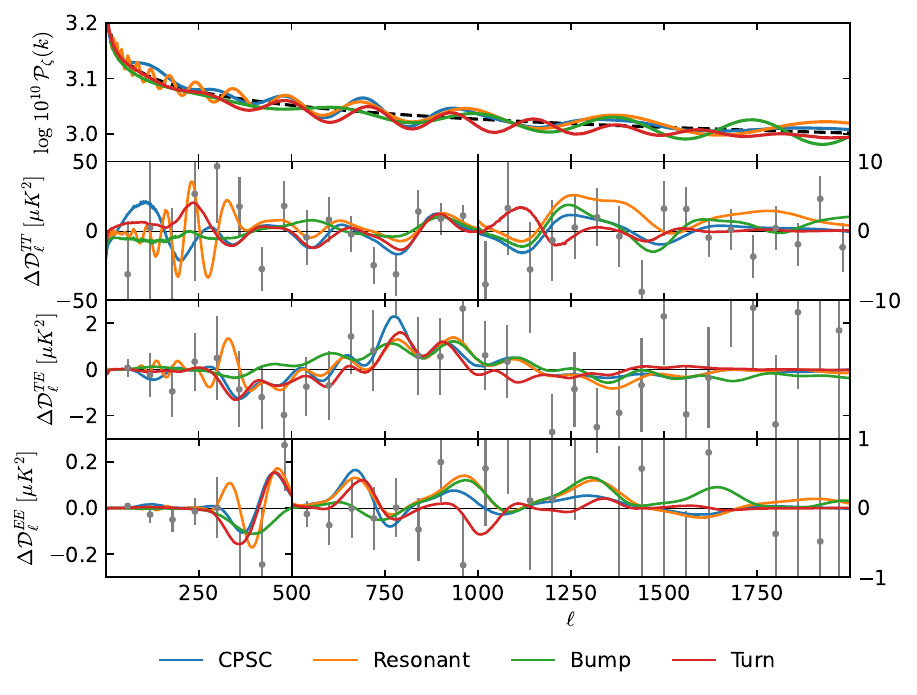}
			\begin{center}\begin{tabular}{|l|l|l|l|l|}
					\hline  
					&  CPSC & Resonant & Bump & Turn\\\hline
					$\ln B$                  & $-1.2\pm0.36$ & $-2.24\pm0.38$  &$-1.44\pm0.36$&$-2.31\pm0.36$ \\         \hline
					$\Delta\chi^2$                &   13.4 & 10.9 & 8.9 & 8.6 \\         \hline
			\end{tabular}\end{center}
			\caption{\footnotesize\label{fig:bestfit_eg20} [Top] Bestfit candidates to Planck data. [Bottom] Bayes factors ($\ln B$) and $\Delta\chi^2=\chi^2_{\rm featureless}-\chi^2_{\rm model}$  obtained for the analysis of Planck data.
			}
		\end{figure*}

The feature models we investigate in this work can be described by the following Lagrangian.
\begin{align}
\mathcal{L} =& -\frac{1}{2} \left[1+ \Xi(\phi) \sigma\right]^2 (\partial\phi)^2
- V_{\rm inf} \left(	1- \frac{1}{2} C_\phi \phi^2  \right)\left(1+\frac{\Delta V}{V}(\phi)\right)\notag\\
\label{eq:Lagrangian}
&-\frac{1}{2} (\partial \sigma)^2- \frac{1}{2} m_\sigma^2 \sigma^2 ~.
\end{align}
The field $\phi$ in the Lagrangian above represents the inflaton, and $V_{\rm inf} \left(	1- \frac{1}{2} C_\phi \phi^2  \right)$ is a small-field slow-roll potential leading to negligible tensor modes.
The field $\sigma$ is a massive field with $m_\sigma\gg H$, we assume it is initially stabilized at $\sigma=0$.
Features in the inflationary trajectory are introduced by non-zero perturbations to either the slow-roll potential, $(\Delta V/V)$, or the kinetic term $\Xi$ of the inflaton.
In the former case, we can simply neglect the dynamics of the massive field $\sigma$, as it does not move from the minimum at $\sigma=0$, therefore making the inflationary dynamics effectively of the single-field kind.
On the other hand, if the coupling $\Xi$ is turned on we have to follow the full two-field dynamics of the Lagrangian~\eqref{eq:Lagrangian}. 

In the following, we will denote the model {\em without} features, i.e. that with $\Xi=\Delta V/V=0$, as the {\rm featureless} model.
Before describing the feature models considered in our paper, we find it useful for the reader to remind the relations between the parameters of the slow-roll potential in Eq.~\eqref{eq:Lagrangian} on one hand, and the tilt and the amplitude of the power spectrum at a given scale $k_0$\footnote{Note that in this paper we compute $A_s$ and $n_s$ at the scale $k_0$ corresponding to the wavenumber that exits the horizon at the time of the feature in the inflationary trajectory, that we will denote as the feature scale, as opposed to the usual definition at the CMB pivot scale $k_*=0.05\,{\rm Mpc}^{-1}$. }
on the other hand:
\begin{equation}
	\label{eq:featureless}
	n_s-1= -2 C_\phi \,\,\,\,\,\,\,\,{\rm and}\,\,\,\,\,\,\,\,\ln\left( 10^{10}\, A_s\right)= \ln\left( 10^{10}\,24\pi^2 \epsilon_0\, V_{\rm inf} \right)
\end{equation}
where the first slow-roll parameter is approximately $\epsilon_0 = \frac{C_\phi^2\,\phi_0^2}{2}$
and background quantities with a subscript $0$ are evaluated at the time implictly defined by the Hubble crossing condition $k_0=a_0 H_0$ for the scale of the feature.
The feature signals studied in this paper are all small scale-dependent corrections to the leading order power-law power spectrum described by the tilt and the amplitude in Eq.~\eqref{eq:featureless}. 

We will be considering 4 representative examples of feature models that we now briefly describe. All the models introduce 3 extra parameters with respect to the SFSR paradigm. This is the minimal number of parameter needed to describe the amplitude, frequency and phase (or envelope, as in some models the phase is uniquely fixed) of the feature.
A schematic representation of the inflationary trajectory for these models is shown in Fig.~\ref{fig:Trajectory} and we plot the corresponding best fit power spectra to Planck data in Fig.~\ref{fig:bestfit_eg20}.  
\begin{itemize}
		\item {\bf CPSC model}: $\Xi =\xi \,{\rm \Theta}(\phi-\phi_0) , \,  \Delta V/V = 0$, where $\Theta$ is the Heaviside step function.
	This is a two-field model of inflation, where the massive field $\sigma$  is excited by a sharp feature and starts to oscillate, before eventually settling down to the minimum of its (effective) potential~\cite{Chen:2014joa, Chen:2014cwa}.
	In this paper, we use the construction of Ref.~\cite{Braglia:2021sun,Braglia:2021rej}, where the sharp feature  is the introduction of a bending curved path of valley, from the time $N_0$ at which $\phi(N_0)=\phi_0$, in the background evolution.
	The correction to the power spectrum consists in a sharp feature signal at large scales, followed by a resonant signal at smaller ones.
	However, unlike in the resonant model (see below) where the oscillations appear at all wavenumbers, the resonant feature in this model is damped and shows up only for a few wavenumbers
	This is easily explained as the background oscillations of the damped field $\sigma$ only last for a finite range of $e$-folds.
	The feature signal is described by 3 parameters: the scale of the feature parameterized by $k_0$ which is the mode crossing the horizon at time $N_0$, the frequency of the resonant signal $m_\sigma/H$ and the maximum amplitude of the signal that can be written in terms of the model parameters as~\cite{Chen:2014cwa,Braglia:2021rej}:
	\begin{equation}
\left.\frac{\Delta \mathcal{P}_{\rm CPSC}}{\mathcal{P}_0}\right\rvert_{\rm maximum}=\sqrt{4\pi}\,\xi^2\epsilon_0\left(\frac{m_\sigma}{H}\right)^{3/2}.
	\end{equation}
	
	The bestfit feature signal to Planck data has an amplitude $\Delta\mathcal{P}/\mathcal{P}_0=0.023$, a frequency  $m_\sigma/H=18.2$ and the sharp feature exciting the massive field oscillations starts at $k_0=2\times10^{-3}\, {\rm Mpc}^{-1}$.

	\item {\bf Resonant model}: 
	$\Xi =0,\,  \Delta V/V = A\, \sin(\phi/\Lambda)$.
	We introduce a periodic modulation to the slow-roll potential, introducing a typical physical frequency $\omega>H$ related to $\Lambda$ and the slow-roll potential~\cite{Chen:2008wn}.
	All physical wavenumbers $k/a$ eventually become comparable to the scale $\omega$, therefore each mode experiences a brief sub-horizon resonance that leads to a characteristic {\em resonant} feature correction to the PPS.
	The discrete shift symmetry in the periodic oscillation of the potential determines the running of the resonant signal to be of the form $\sin\left(\frac{\omega}{H} \ln k\, {\rm Mpc}+ {\rm phase} \right)$, with an approximate constant amplitude. So, only for this simple model we simply use the following template for the PPS in our analysis, rather than resorting to a full numerical resolution of the inflationary dynamics from the potential: 
	\begin{equation}
\frac{		\Delta \mathcal{P}_{\rm res}(k)}{\mathcal{P}_0}=\frac{		\Delta \mathcal{P}}{\mathcal{P}_0} \sin\left( \frac{\omega}{H}\ln k \, {\rm Mpc}+ \varphi \right).
	\end{equation} 
	
		The bestfit feature signal to Planck data has an amplitude $\Delta\mathcal{P}/\mathcal{P}_0=0.015$, a frequency  $\omega/H=18.1$ and a phase $\varphi=5.9$. It is worth noting that the bestfit frequency of the signal is the same as in the CPSC bestfit. In fact, the two spectra overlap in the region   $0.03\,{\rm Mpc}^{-1}\,\lesssim k \lesssim 0.07 \,{\rm Mpc}^{-1}\,$ and fit similar residuals at the multipoles $400\lesssim\ell\lesssim1000$, providing a better improvement to the fit compared to the sharp feature models that we are going to introduce. Taken at face value, this is a possible hint at a mechanism active during inflation that produces a resonant signal with a characteristic frequency of $\omega/H\sim18$ at the scales of $400\lesssim\ell\lesssim1000$.

	\item {\bf Bump or dip model}: $\Xi = 0,\,  \Delta V/V = A\, \exp\left[-\left(\phi-\phi_0\right)^2/\Delta^2\right]$.  We introduce a sharp feature with a Gaussian profile in the inflaton potential, centered around the inflaton value $\phi_0$ and with a typical field-space width of $\Delta$~\cite{WMAP:2003syu}.
	Note that the profile of the feature is crucial as the signal is very sensitive to it.
	For example, popular choices also include step~\cite{Adams:2001vc} or kink-like features~\cite{Starobinsky:1992ts}.
	These models have been studied extensively in the literature, but they result in features that can fit well low-$\ell$ anomalies, but not high-$\ell$ ones, which are the focus of this paper. Models that can address both large and small-scale anomalies do exist, but require more than 3 extra  parameters~\cite{Hazra:2014goa,Braglia:2021sun,Braglia:2021rej,Hazra:2021eqk}.
	As other sharp features, the bump model results in oscillations linear in $k$-space.
	It can be described by 3 parameters: the mode $k_0$ crossing the horizon at the time of the sharp feature $N_0$, the duration inflaton spends crossing the feature $\Delta N\simeq6\Delta/\sqrt{2\epsilon_0}$ and the maximum feature amplitude that is found as (see App.~\ref{app:analytical_results}):
	\begin{equation}
		\left.\frac{		\Delta \mathcal{P}_{\rm bump}}{\mathcal{P}_0}\right\rvert_{\rm maximum}=\frac{3 A}{\epsilon_0}\sqrt{\frac{2\pi}{e}}.
	\end{equation}
Note that the sign of the parameter $A$ in the potential is free (the feature can be either a bump or a dip), so the effective parameter $\Delta \mathcal{P}_{\rm bump}/\mathcal{P}_0$ can take both positive and negative values, effectively amounting to a phase shift of $\pi$ in the oscillations. 

	The bestfit feature signal to Planck data has an amplitude $\Delta\mathcal{P}/P_0=0.04$, a frequency  $k_0=4.1\times10^{-3}\, {\rm Mpc}^{-1}$ and a very short duration $\Delta N =0.09$ $e$-folds (note the positive sign of the amplitude which means that the feature is due to a bump, not a dip). Interestingly, this is the candidate with the largest amplitude. However, due to the relatively large $k_0$, the amplitude of the feature signal in $\mathcal{P}(k)$ is growing in the observable range and peaks at very small scales to which Planck is not sensitive.  Note, however, that since the sensitivity of  CMB angular spectra to sharp features decreases with $\ell$ (see Appendix~\ref{app:cmb_estimates}), this compensates with the growing amplitude and the CMB residuals do not grow towards high multipoles. We note that, while the multimodal nature of the (unconstrained) feature parameters is expected, the reader might be surprised to see an apparent bimodality of the overall amplitude $\ln 10^{10}\,A_s$, that presents two peaks at $\ln 10^{10}\,A_s^A\sim3.135$ and $\ln 10^{10}\,A_s^B\sim3.057$, the latter being close to the bestfit value. In fact, this does not mean at all that the amplitude is not constrained. Rather, it is a consequence of our choice of computing $\ln 10^{10}\,A_s$ at the feature scale $k_0$, instead of a fixed reference scale $k_*$, as done more often. While this is convenient for all the other models, it leads to a slightly bimodal posterior for $\ln 10^{10}\,A_s$ in the bump model, in proximity of two different values of $N_0-N_*$. The bump/dip model attempts to fit the low-$\ell$ dip around $\ell\sim20-30$ in the Planck temperature data, which is also clear from the bottom-left panel of Fig.~6 in Section~\ref{sec:features}. Therefore $k_0^A\ll k_0^B$ and $N_0^A< N_0^B$, with the mode $B$ being the bestfit candidate that fits high-$\ell$ residuals. To maintain the overall amplitude of $\mathcal{P}(k)$ across a wide range of scales, we must have $\ln 10^{10}\,A_s^A>\ln 10^{10}\,A_s^B$, hence the apparent bi-modality.

	\item {\bf Turn model}: $\Xi = \xi\exp\left[-\left(\phi-\phi_0\right)^2/\Delta^2\right], \, \Delta V/V = 0$.
	We introduce a localized kinetic coupling between the inflaton and the massive field, which is stabilized at $\sigma=0$. The coupling has a Gaussian profile and is switched on around the time $N_0$, where we used the same notations as before. This model is phenomenologically similar, although not exactly the same, as the turn model in Ref.~\cite{Achucarro:2010da}.
	We also note that the difference between the turn model and CPSC model is that, in the former case, the coupling is turned on only for a very short time leading to a purely sharp feature, with no resonant signal imprinted, because the $\sigma$ field is either not oscillating, or oscillating along a straight trajectory without direct coupling to the inflaton so that the resonant signal is too small to show up \cite{Gao:2012uq,Gao:2013ota}.
	In terms of field-space trajectory, while the CPSC is described by a straight line followed by a bent trajectory, here the trajectory simply consists in two straight lines joined by a sharp turn, as can be seen from Fig.~\ref{fig:Trajectory}.
	As shown in App.~\ref{app:analytical_results}, the feature signal is very similar to the dip one, except for the first oscillation at the largest scales which is a bit more pronounced in this model.
	For this reason, it is instructive to consider this model to investigate whether future experiments will be sufficiently accurate such as to distinguish it from the dip one.
	The turn model can be described by $k_0$, the duration of the feature $\Delta N= 6 \Delta /\sqrt{2\epsilon_0}$ and the maximum amplitude (see App.~\ref{app:analytical_results}):
			\begin{equation}
			\left.\frac{		\Delta \mathcal{P}_{\rm turn}}{\mathcal{P}_0}\right\rvert_{\rm maximum}=2\epsilon_0\left(\frac{\xi}{m_\sigma/H}\right)^2\sqrt{\frac{2\pi}{e}}.
		\end{equation}
	Note that, as opposed to the bump feature, here $	\Delta \mathcal{P}_{\rm turn}/\mathcal{P}_0$ can only take positive values, since it is quadratically dependent on the coupling constant $\xi$. For this reason, the model can only be similar to half of the feature signals produced by the bump/dip model, i.e. those from a dip in the potential.

	The bestfit feature signal to Planck data has an amplitude $\Delta\mathcal{P}/P_0=0.02$, a frequency  $k_0=2.5\times10^{-3}\, {\rm Mpc}^{-1}$ and duration $\Delta N =0.58$ $e$-folds, so the feature is not as sharp as for the bump case.

\end{itemize}
 
We note that feature parameters are not constrained\footnote{For constraints on the feature parameters from Planck data see triangle plots in the next Sections.}, consistently with the Bayes factors reported in the Table in Fig.~\ref{fig:bestfit_eg20}, that show no evidence for features. Nevertheless, Planck data do set stringent bounds on the amplitude of such features. While depending quantitatively on the model at hand, Planck bounds on the feature amplitude are generically $\Delta\mathcal{P}/\mathcal{P}_0<\mathcal{O}(0.05)$ (see next Sections and Ref.~\cite{Planck:2018jri}).
As can be seen from Fig.~\ref{fig:bestfit_eg20}, the best-fit candidates for each model address the oscillatory residuals in the high-$\ell$ temperature Planck data. While the $\chi^2$ are different for each best fit, we see from the Bayes factors that none of the models is statistically favored over the featureless case.
This can be explained by the fact that they require extra parameters, which penalizes their Bayesian evidence.
It is also interesting to see that the correlated features in the $TE$ and $EE$ spectra are very different depending on the model, but that we can only partially tap into such information because the error bars are much larger than those in the $TT$ spectrum.
This motivates the analysis of the next Sections, where we show that shrinking such error bars will indeed open up a window on distinguishing feature models one from the others, and assessing their statistical significance.

As mentioned in the Introduction, our forecast analysis is based 
on picking the best-fit candidate of these models, one model after the other, and assuming it as a fiducial model of the universe.
Then, other models are tested against this fiducial universe by performing a nested sampling. In particular, we will be using three combinations of present and future datasets, namely, Planck + Simons Observatory (PL + SO), Planck + LiteBIRD + Simons Observatory (PL + LB + SO) and Planck + LiteBIRD + CMB Stage-4 (PL + LB + S4)). We provide more details on our forecast analysis in Appendix~\ref{app:analysis}. For an in-depth discussion of all the technical details, as well as the noise specifications of the experiments we use, we refer the reader to~\cite{Braglia:2021rej}, which we follow closely.
If we wanted to exhaust all possible combinations of the 4 feature models, plus the featureless one, we would have to perform 25 separate analysis, which would unnecessarily complicate the understanding of the physical results presented in this work.
Therefore, we decided to perform only some representative comparisons that highlight a specific information that we will get access to with future CMB experiments.
We will provide more details in each of the next Sections.

\section{Ruling out features {\em if} the Universe is featureless}
\label{sec:featureless}
\begin{figure}
	\includegraphics[width=.43\columnwidth]{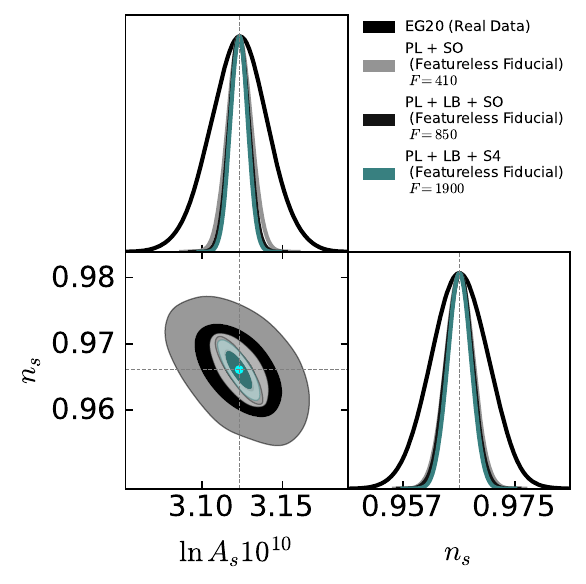}
	\includegraphics[width=.57\columnwidth]{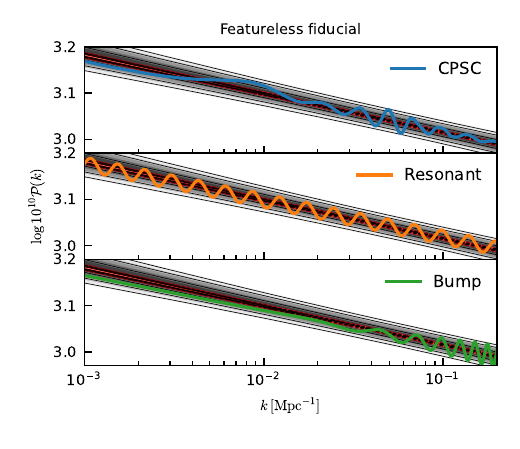}
	\caption{\footnotesize\label{fig:featureless} [Left] Constraints on $A_s$ and $n_s$ when the featureless bestfit model is assumed as fiducial.
	We mark the Planck bestfit with a small circle. In the legend we provide the values for the gain in the ${\rm FoM}$ in going from one survey to another.
	[Right] Reconstructed signal from Planck data (gray colormap) and from our forecast analysis for PL + LB + S4 (red colormap). We plot different shades in going from 0 (inner regions, darker shades) to 3$\sigma$ (outer regions, lighter shades). On top of the reconstructed signal, we plot the current feature best fits to show that they are consistent with Planck data, but also that they will be all excluded by future measurements if the featureless fiducial is the true model of the Universe.}
\end{figure}

\begin{figure}
	\begin{center}
		\includegraphics[width=.495\columnwidth]{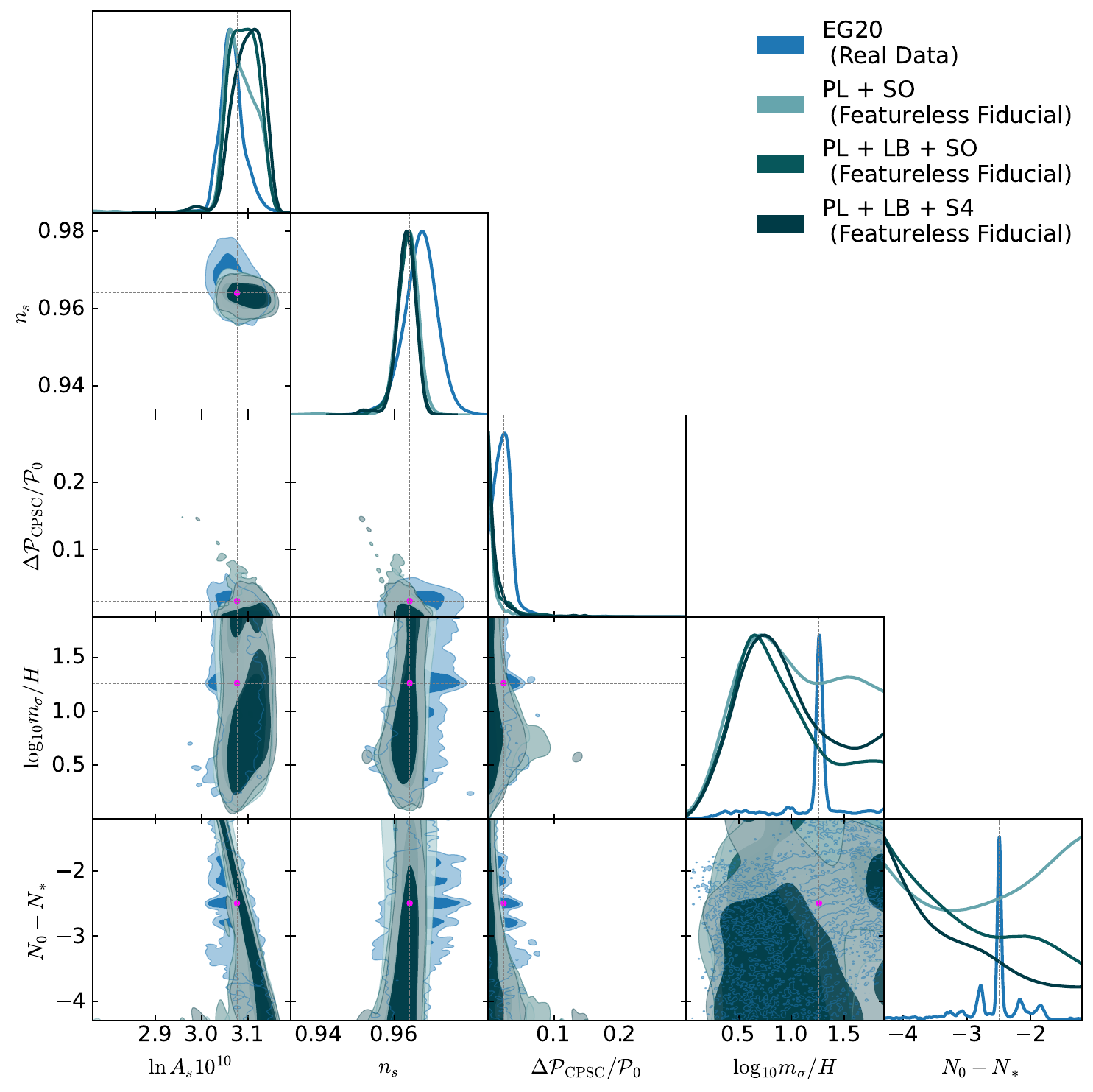}
		\includegraphics[width=.495\columnwidth]{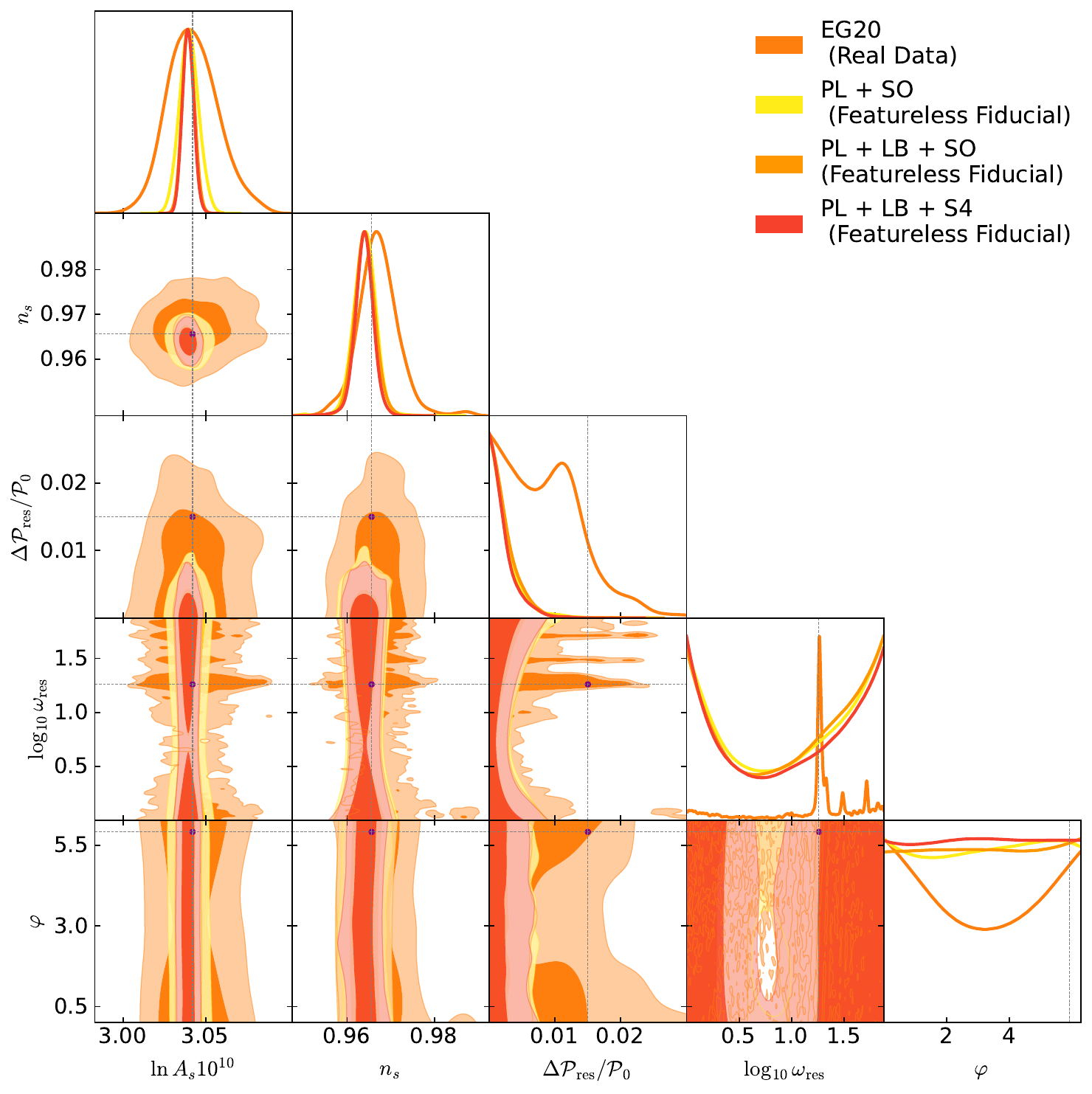}
		\includegraphics[width=.495\columnwidth]{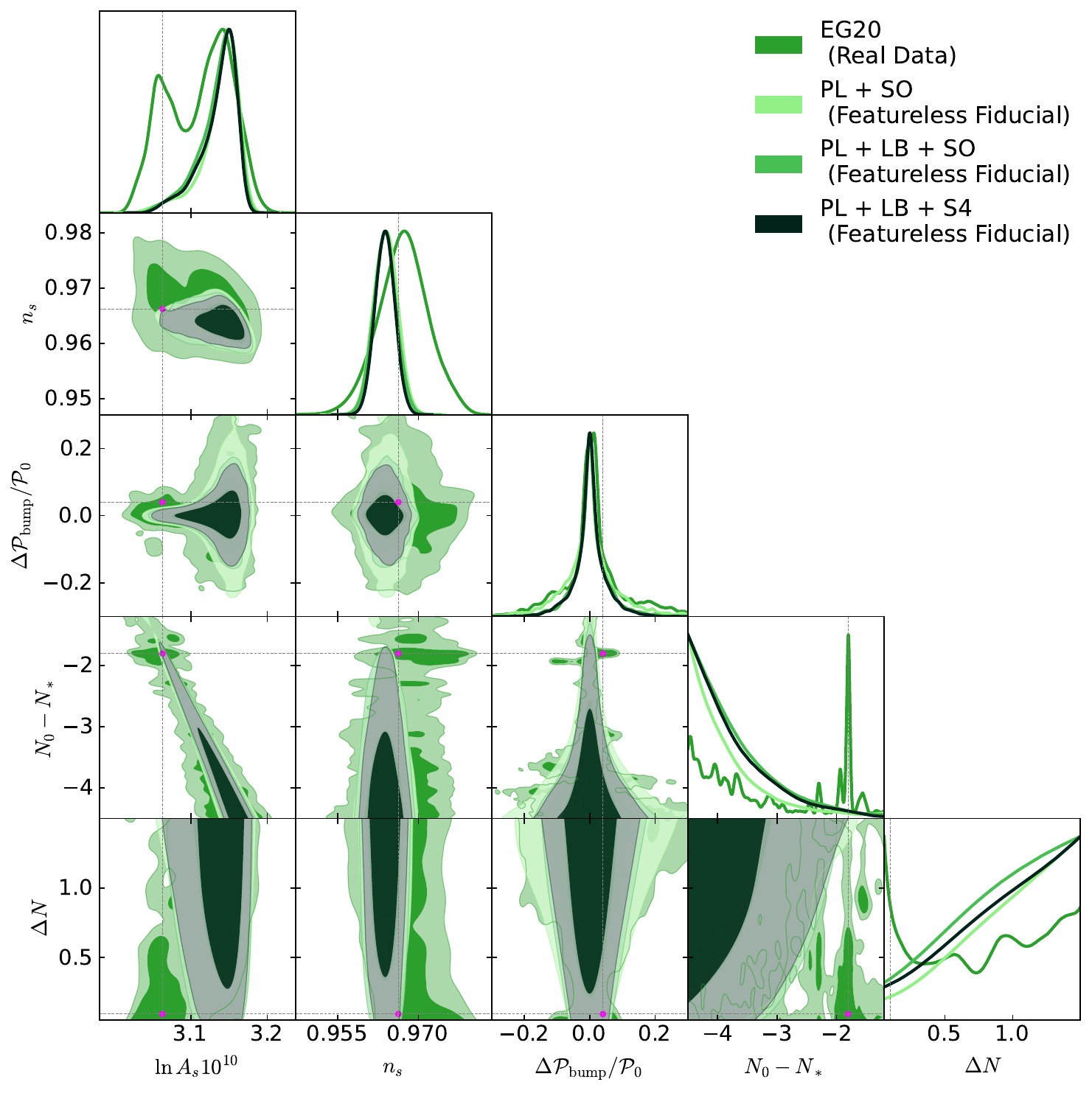}
		\includegraphics[width=.48\columnwidth]{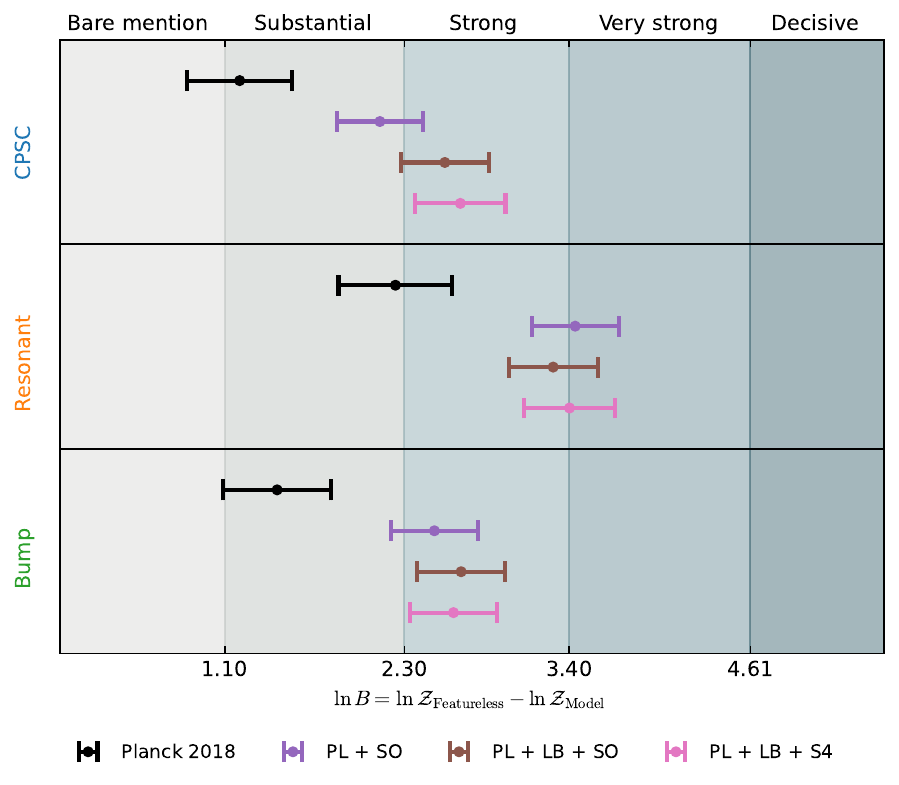}\end{center}
	\caption{\footnotesize\label{fig:featureless_comparison} 
	Constraints on the CPSC (top-left), resonant (top-right) and bump (bottom-left) model.   We assume the featureless bestfit model as fiducial. We mark the Planck bestfit of each model with small circles. Projected Bayes factors ($\ln B$) obtained for different experiments {\it w.r.t.}~the Featureless bestfit model, which is assumed as fiducial cosmology (bottom-right).	}
\end{figure}

We start our analysis by addressing the most simple question, which is also the most conservative one given the success of SFSR models.
This question is the following: would be able to rule out features in case they are just statistical fluctuations?
We therefore assume the featureless best fit as our fiducial and compare our features model to it.
We stress, however, that presently feature models are only slightly disfavored with respect to the featureless one, so, in the Bayesian interpretation, this hypothesis has no special status. In order to see the power of future experiments to rule out each of the feature best fit candidates, we perform the analysis against the fiducial for all of the models but the Turn one, since it is very similar to the Dip one and therefore we do not expect significant differences.

Let us first discuss the constraints on the featureless power spectrum that we will get from future experiments. These are shown in Fig.~\ref{fig:featureless}. The improved error bars on the E-mode measurements will significantly shrink the posterior with an improvement of the FoM of a factor of $\mathcal{O}(10^2)$ to  $\mathcal{O}(10^3)$, with the better improvement which naturally comes by adding both large scale and small scale measurements from LiteBird and S4. In particular, the constraint on the spectral index $n_s = 0.9661\pm 0.0019$ at 68 \% CL from PL + LB + S4 will improve current constraints by roughly a factor of two, setting under more pressure some of the popular single-field slow-roll models\footnote{Note that the fiducial value of $n_s$ that we adopt in this Section ($n_s=0.9661$) is slightly different from the one quoted in the Introduction ($n_s=0.9649$). The reason is that the former is the fiducial from the analysis with the EG20 likelihood~\cite{Efstathiou:2019mdh} (adopted in this work), while the latter comes from the analysis in the official Planck paper~\cite{Planck:2018jri}. The two values are consistent for current errors on $n_s$ and this does not affect our analysis. }. 

In the right panel of Fig.~\ref{fig:featureless}, we also show the so-called predictive posterior distribution on the power spectrum on which we overplot the feature best fit candidates to Planck data\footnote{We use the code fgivenx~\cite{fgivenx}, which was also used for plotting the functional posterior distribution of feature models in~\cite{Hergt:2018ksk} and for free form reconstructions in~\cite{Handley:2019fll}. }. This clearly illustrates why those candidates, although being perfectly consistent with current data, will be completely excluded if the featureless model is really the one describing our Universe. Indeed, all the candidates fall outside the red band corresponding to the 1 to 3$\sigma$ constraint on the featureless $\mathcal{P}(k)$.

We show in Fig.~\ref{fig:featureless_comparison} the constraints on each feature model for each of the 3 experimental setups considered. All the contour plots share the same characteristics. First of all, the parameter controlling the amplitude of the feature is bounded more with respect to its current constraint. Second, all the modes in the frequency or location parameters are washed out and the corresponding forecast posteriors are almost flat, signaling that we will not get any hints for features. From the Bayes factor analysis (see bottom-right panel), the evidence against features will be generally strong, but not decisive. This seemingly curious fact is instead rather easy to understand.
While feature models will be put under strong pressure, and we will rule out all current best fits to Planck data, features with small amplitude will still be consistent with future data.
So although the Bayesian evidence will be penalized by the number of extra parameters, the fit will always be slightly better or equal to the one from the featureless model, never leading to a decisive evidence against features.

\section{Detecting features {\em if } they are the true model of the Universe }
\label{sec:features}			
			
\begin{figure}
					\includegraphics[width=.5\columnwidth]{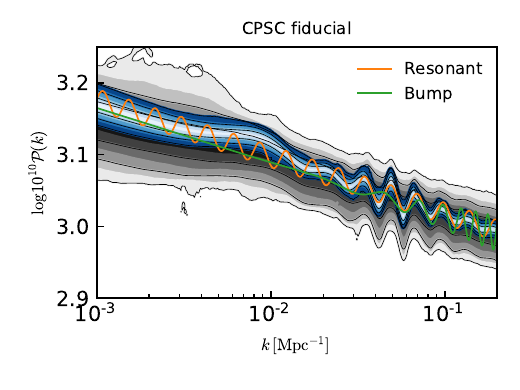}
					\includegraphics[width=.5\columnwidth]{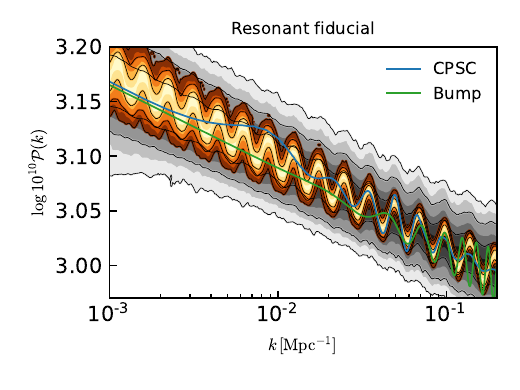}
					\includegraphics[width=.5\columnwidth]{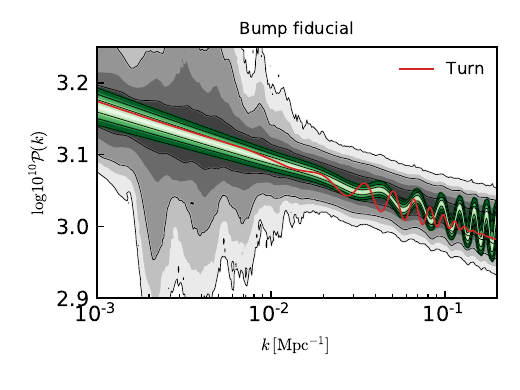}
					\includegraphics[width=.5\columnwidth]{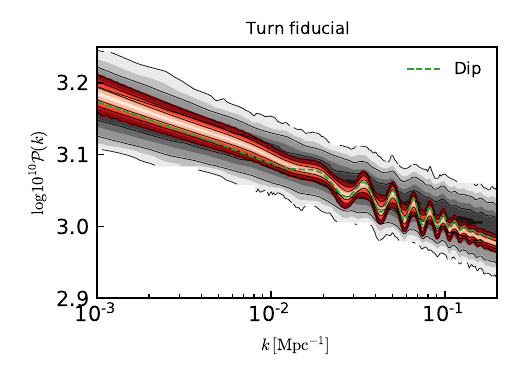}
					\caption{\footnotesize\label{fig:reconstructions}  Reconstructed spectra for our feature models assuming in each case that their best fits to P18 represent the true model of the Universe. Gray and red-shaded regions correspond to 0 to 3$\sigma$ reconstructed CL from Planck data and  PL + LB + S4 forecasts respectively.
				}
					
				\end{figure}

We now take a further step towards addressing the main question of this paper, i.e. whether we will be able to distinguish different feature models using future observations of the CMB polarization.
In this short Section, we show the evolution of the constraints on the power spectrum of our models going from current data to forecasts for future experiments. This step will be very helpful to interpret the results of the next Sections.

Technically, we run our nested sampling forecast for each feature model assuming the corresponding Planck best fit, presented in Section~\ref{sec:models}, as fiducial. Then, we plot the resulting predictive posterior distribution for $\mathcal{P}(k)$. The constraints on the model parameters obtained from our runs are shown in the next Sections. The results for each of our 4 feature models are shown in Fig.~\ref{fig:reconstructions}, where, for simplicity, we only plot the predictive posterior from our PL + LB + S4 analysis (blue, orange, green and red contours for CPSC, resonant, bump and turn models respectively) on top of that from real Planck data (gray contours)\footnote{Note that the reconstructed contours from Planck in the case of the bump model show hints for the dip anomaly around $\ell\sim20-30$, which are not present in the case of the turn, which has a different phase of the feature signal.}.

At this stage of our paper, the idea that we would like to convey is that the constraints on the primordial features will become extremely tight in the future, to the point that candidates from other feature models fall outside the 3$\sigma$ posterior and thus will likely be ruled out.  
The shrinking of the posterior is accompanied by an increase in the FoM which ranges from $\mathcal{O}(10^5)$ to $\mathcal{O}(10^9)$ depending on the model (see next Sections for the precise values), which reflects the compression of the parameter space with respect to the analysis with real Planck data. We also note that for each analysis, the featureless spectrum is ruled out with very strong or decisive evidence depending on the model considered.
We now move to a more rigorous analysis, confirming all the statements we made in this Section.

			\section{Discriminating between feature models}
			\label{sec:comparison1}

\begin{figure}
			\begin{center}	\includegraphics[width=.45\columnwidth]{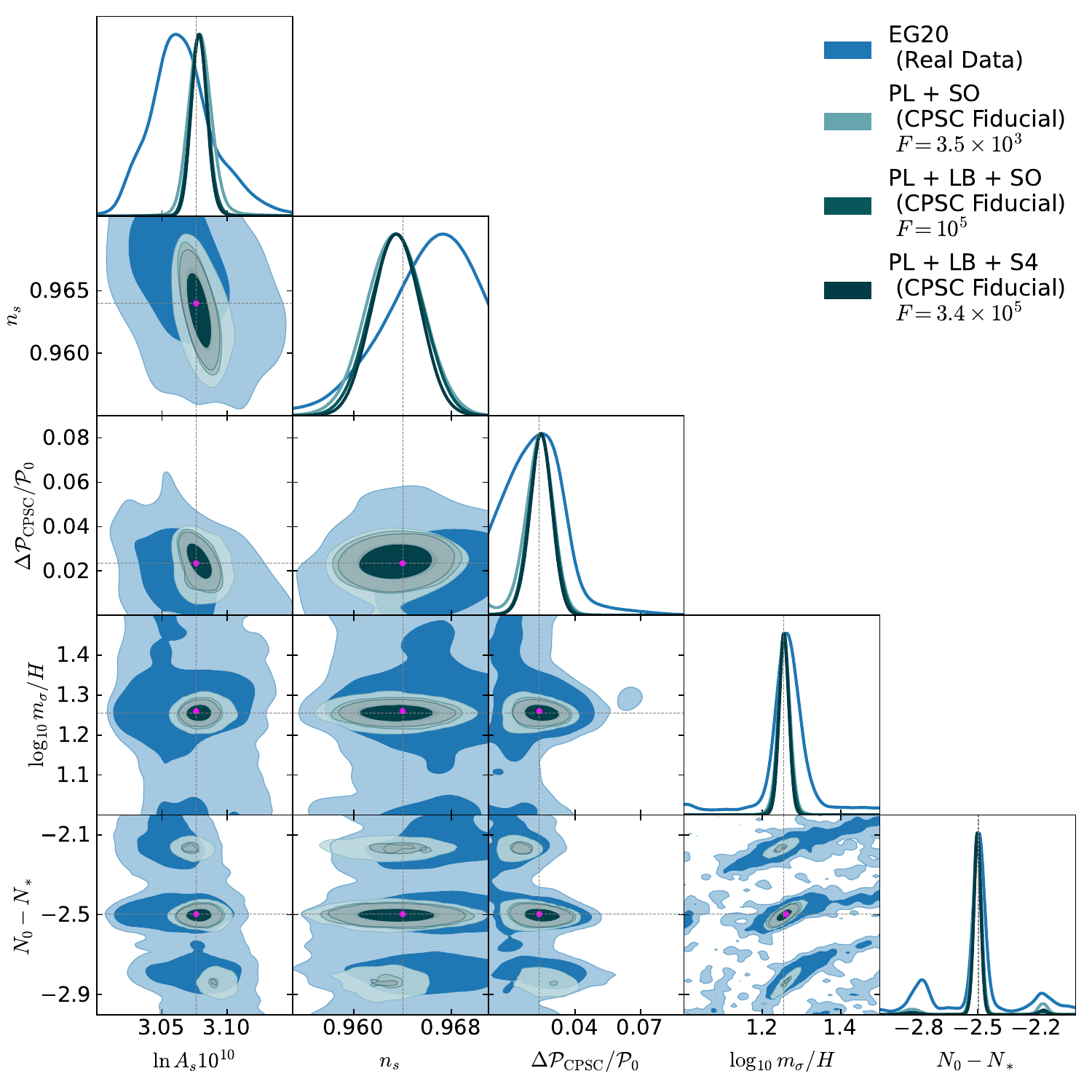}
		\includegraphics[width=.45\columnwidth]{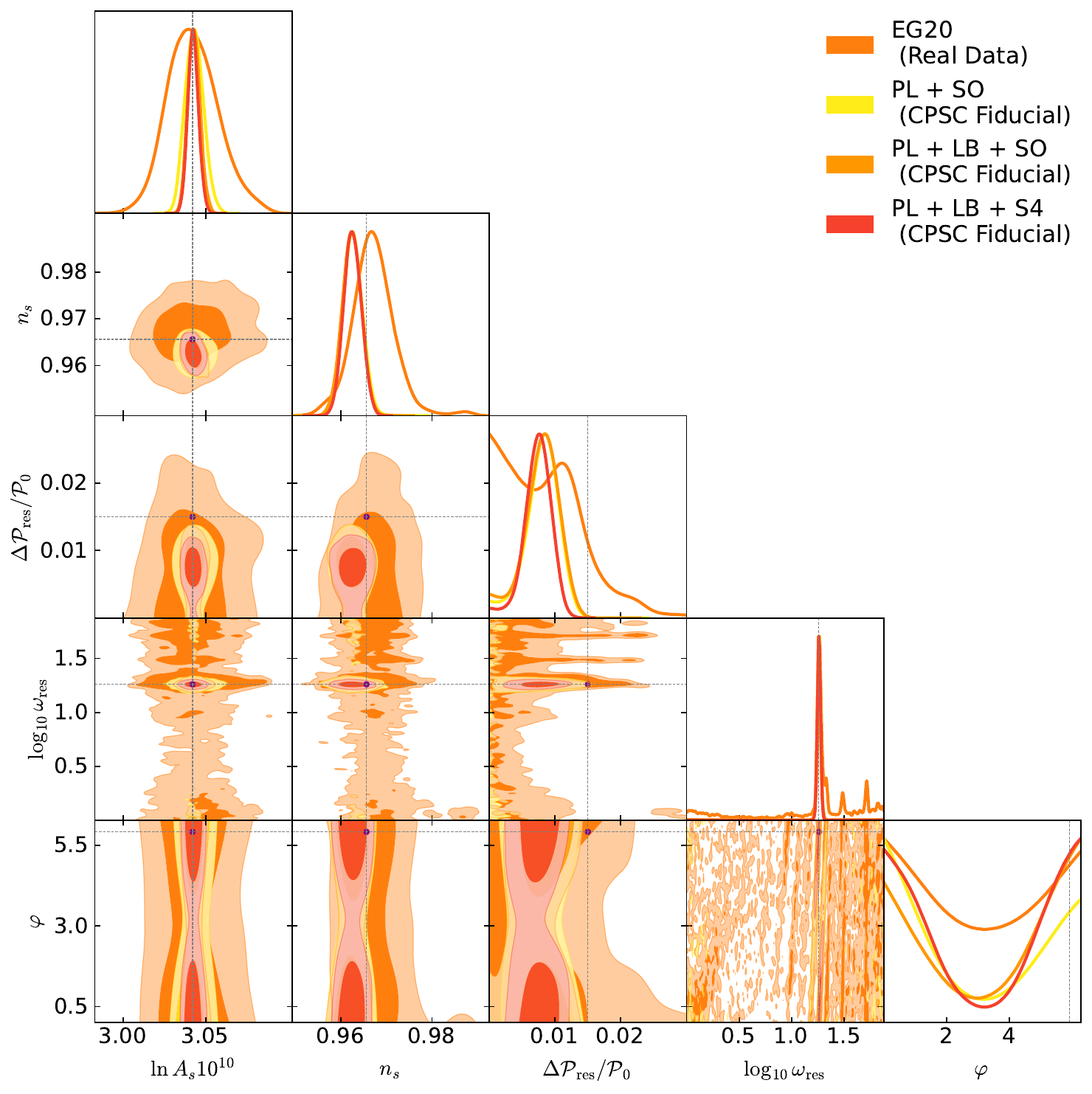}			\includegraphics[width=.45\columnwidth]{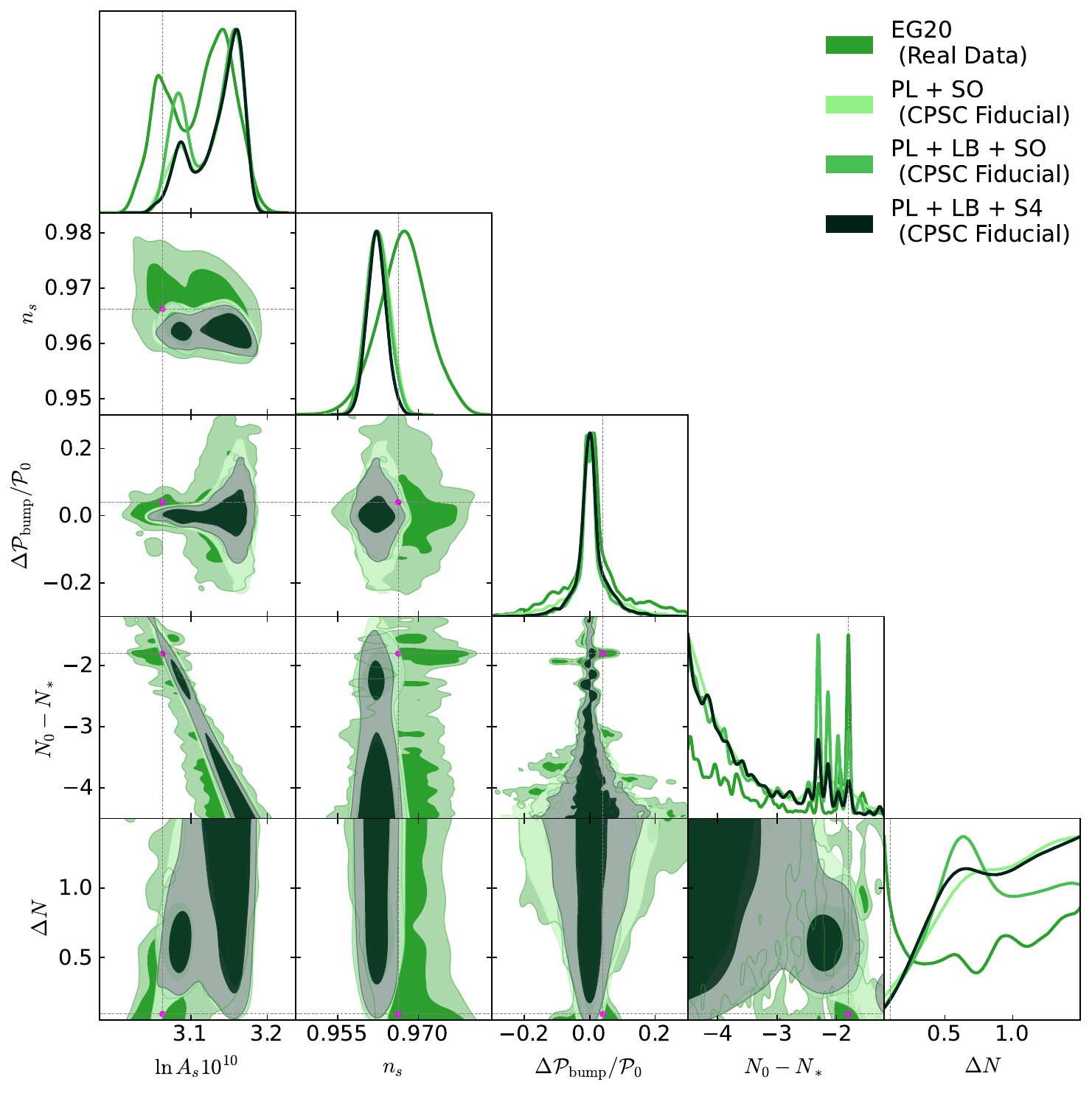}
				\includegraphics[width=.45\columnwidth]{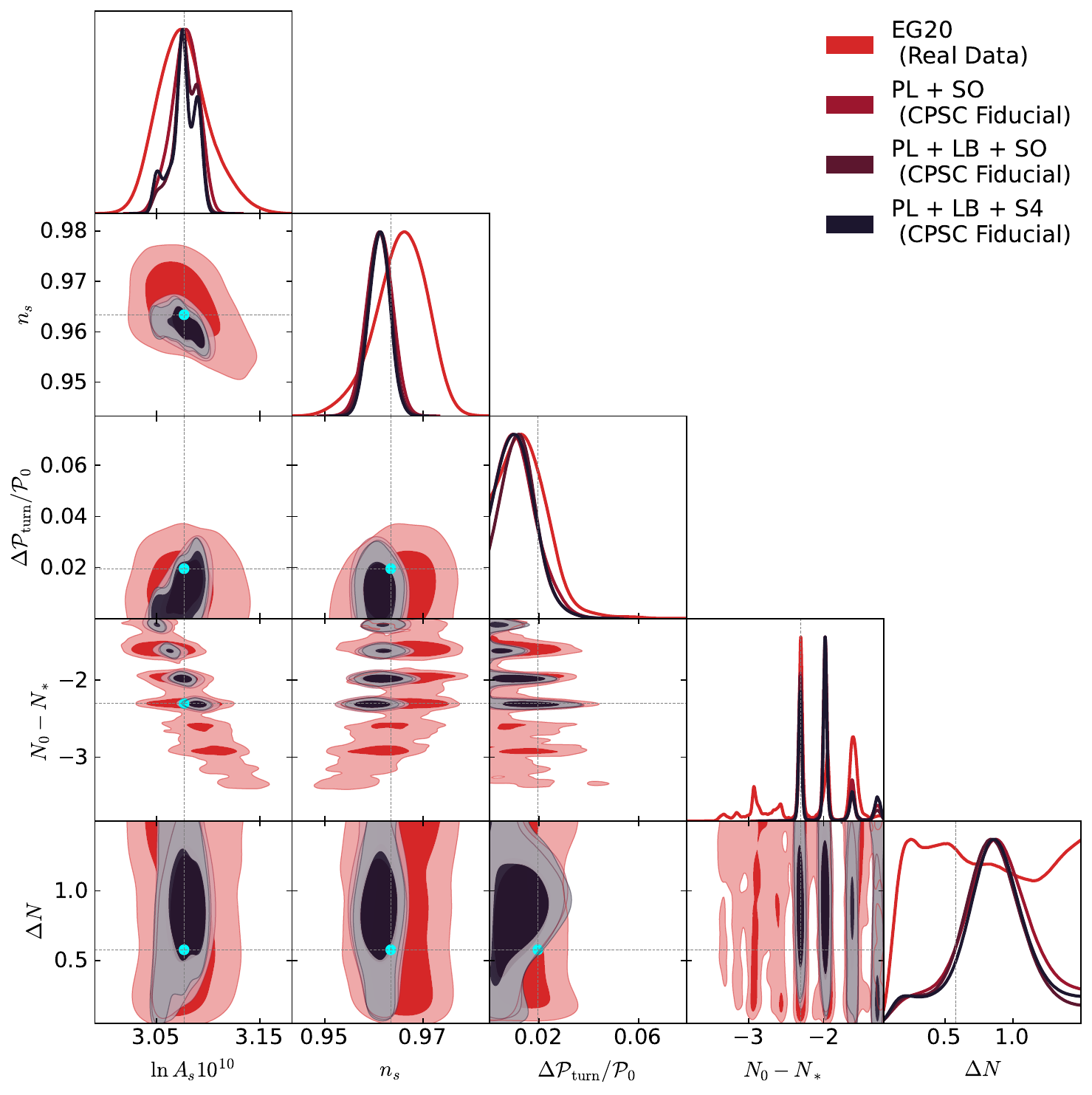}
				\includegraphics[width=.4\columnwidth]{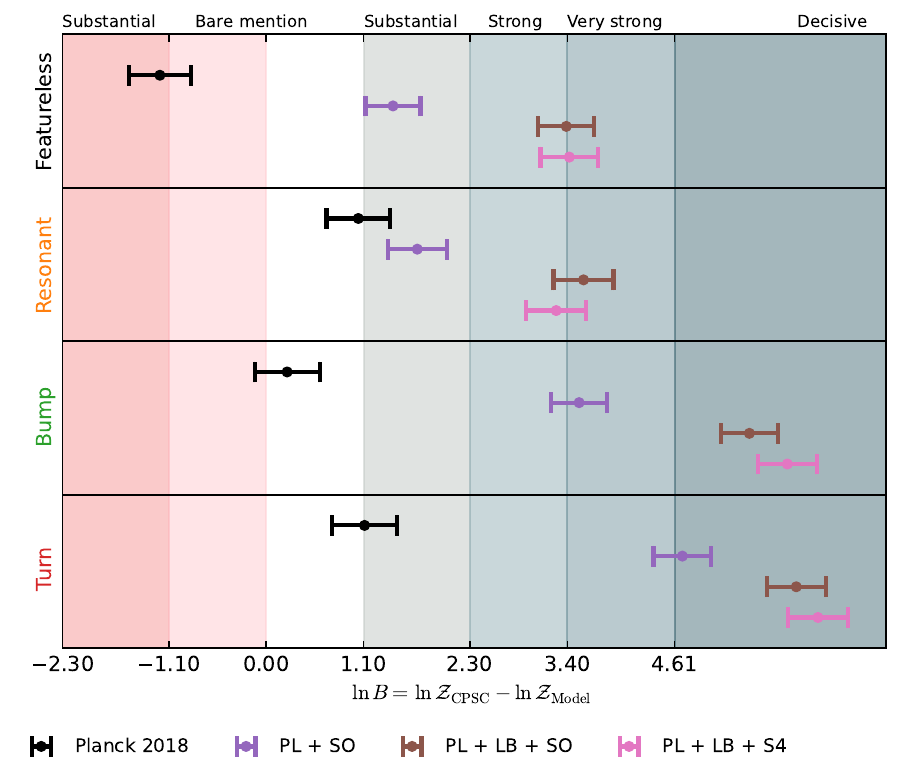}
			\end{center}	\caption{\footnotesize\label{fig:cpsc_comparison} 
					Constraints on the CPSC(top-left), resonant (top-right), bump/dip (bottom-left) and turn (bottom-right) model.   We assume the CPSC bestfit model as fiducial. We mark the Planck bestfit of each model with small circles. In the bottom-most panel we plot the projected Bayes factors ($\ln B$) obtained for different experiments {\it w.r.t.}~the fiducial CPSC bestfit model.}
				
			\end{figure}

We now assume the CPSC best fit and compare to it the other feature models. First of all, let us show in the top-left panel of Fig.~\ref{fig:cpsc_comparison}  the projected constraints on the CPSC model itself. Future experiments will be able to identify, out of the various peaks of the multimodal posterior distribution from Planck data, the main peak; the secondary ones will be washed out.
The constraints around the fiducial point will be extremely tight, with a significant increase of the FoM up to $3.4\times10^5$ for PL + LB + S4. In particular, we will have a clear detection of the amplitude $\Delta \mathcal{P}_{\rm CPSC}/\mathcal{P}_0$.
We see that measurements at high-$\ell$ from the SO will already provide a clear-cut hint to a non-zero feature amplitude within the next decade. The addition of large-scale data from LB will push the detection level beyond 4$\sigma$, leading to a constraint $\Delta \mathcal{P}_{\rm CPSC}/\mathcal{P}_0 = 0.0241\pm 0.0057$, i.e. a clear detection at 4.2$\sigma$, which will be further tightened by S4 to $\Delta \mathcal{P}_{\rm CPSC}/\mathcal{P}_0 = 0.0241\pm 0.0053$, reaching the 4.5$\sigma$ detection~\cite{Braglia:2021rej}.
Furthermore, once the amplitude $\Delta \mathcal{P}_{\rm CPSC}/\mathcal{P}_0$ is detected, we will be able to identify the frequency of the clock signal with extreme accuracy for $m_\sigma/H=17.8^{+1.9}_{-1.5}$ for PL + SO, $m_\sigma/H=18.0\pm 1.4$  PL + LB + SO and  $m_\sigma/H=18.0\pm 1.1$  PL + LB + S4 respectively\footnote{Note that these results are slightly different from those obtained in Ref.~\cite{Braglia:2021rej}. 
This is because the models considered are not exactly the same. The one in Ref.~\cite{Braglia:2021rej} was the particular limit of a more sophisticated model able to fit also large scale anomalies. While the best fit of the model match to a very high accuracy, the mapping of the effective amplitude parameter to its exact numerical amplitude only worked as an order of magnitude estimate in Ref.~\cite{Braglia:2021rej}, while it is almost perfect here, leading to a small difference in the constraints derived in the two papers.
Furthermore, the location of the sharp feature is parameterized in a different way in the two papers.}.

If CPSC can be distinguished from other feature models, these results will amount to a clear detection of a massive particle with its mass precisely measured relative to the Hubble rate and provide a direct evidence for the phase of inflation responsible for the production of the anisotropies in the CMB. 
However, while it is clear that the featureless form of the power spectrum will be excluded if the CPSC best fit is really the model of the Universe, it is less clear whether we could mistake other primordial feature signals for the CPSC best fit.
In fact, this is a general question that would have to be asked in the event of a detection of any kind of features in the density perturbations -- to what extent are we able to distinguish different feature models?

To answer this question in our context, we thus compare our feature models to the fiducial one (the CPSC current bestfit) and check whether the resulting posterior distributions hint at any other model candidate and if the Bayesian evidence shows preference for these models.
We show the results of our analysis in Fig.~\ref{fig:cpsc_comparison}. Clearly, both the Bump and Turn models (bottom panels, respectively left and right) show no constraints on the feature parameters, except for a tighter constraint on the feature amplitude {\it w.r.t.} the one from Planck data.
On the other hand, it is interesting to see that the Resonant model (upper right panel) does show a peak in the posterior distribution corresponding to the frequency of the clock signal in the fiducial model.
The peak in  the posterior for the amplitude however is consistent with $0$ at 2$\sigma$, if we approximate the slightly skewed posterior on $\Delta \mathcal{P}_{\rm res}/\mathcal{P}$ to a Gaussian.
This can be explained as, although the model decently fits the first 2 oscillations of the clock signal, its signal has a constant amplitude which does not fit well the resonant part of the CPSC fiducial, whose oscillations are instead damped.
Furthermore, the Resonant model is not able to fit the sharp feature part of the CPSC best fit.
The plot in the bottom-most panel of Fig.~\ref{fig:cpsc_comparison} shows that there would be mounting evidence in favor of the CPSC best fit, consistent with what we have just discussed.
We thus conclude that we will be able to rule out sharp features as the origin of the feature signal and put under very strong pressure a purely resonant signal, if the true model of the Universe is the CPSC best fit to the current data.

In Sec.~\ref{sec:estimator} we will show that, globally over all scales, models used in the above analyses are all $40\%$ to $80\%$ correlated with the fiducial CPSC model.
Despite these correlations, we see that we will still be able to distinguish them using CMB alone due to the rich information feature signals carry in their phases and envelops across different scales. 
So our analysis shows the considerable discriminating power of upcoming CMB experiments for pinpointing the physical mechanism during inflation responsible for the feature signal.
We expect such a prospect to become even more promising when information of other tracers of density perturbations, such as galaxies, cosmic hydrogen atoms and stochastic gravitational waves are added.

						\begin{figure}
							\begin{center}
	\includegraphics[width=.4\columnwidth]{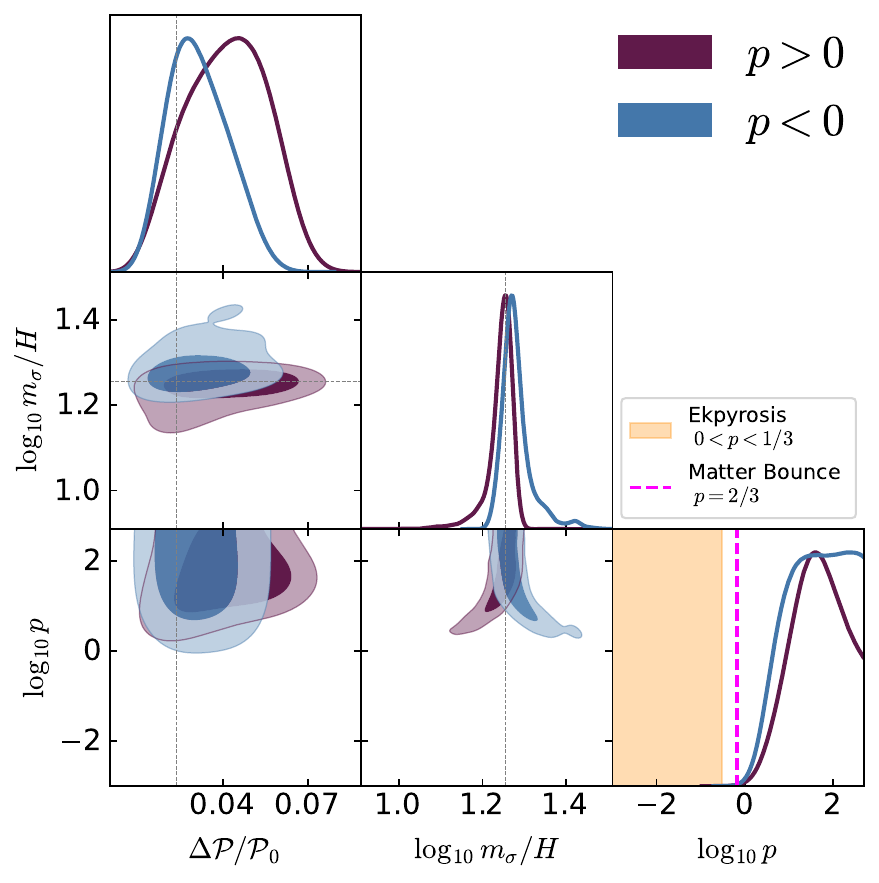}
	\caption{\footnotesize\label{fig:alternative_scenarios}   Constraints on the general clock template from a forecast for PL+LB+S4 assuming the CPSC best fit.  
	}
\end{center}	
\end{figure}

{\textbf{Ruling out alternative scenarios to inflation.} A very interesting property of CPSC models is that the clock part of the signal encodes information about the time-dependence of the scale factor itself, $a(t)\sim t^p$, during the early Universe evolution~\cite{Chen:2011zf}. The signal can be described in a model independent way by the following template:
\begin{equation}
	\frac{\Delta \mathcal{P}_{\rm clock}}{\mathcal{P}_0} =
			A \left( \frac{2k}{k_r}  \right)^{-\frac{3}{2}+\frac{1}{2p}}
			\sin \left[ p\,\frac{m_\sigma}{H} \left( \frac{2k}{k_r}  \right)^{\frac{1}{p} } + \varphi \right], 
	\label{eq:clock_template}
\end{equation}
where $k_r$ is the first resonant mode and we see that the running of the clock signal, i.e. the functional form of the argument of the oscillation, is essentially the inverse function of the scale factor: $t(a)\sim a^{1/p}$ which is $ \sim k^{1/p}$ at horizon crossing for the mode $k$. The above template is valid for $k<2\, k_r$ for $0<p<1$ and for $k>2\, k_r$ otherwise. We note that inflation corresponds to the large $\lvert p\vert$ limit, where  $p (m_\sigma/H) (2k/k_r)^{1/p}\rightarrow \delta\varphi + (m_\sigma/H)\ln(2k/k_r)$, where $\delta\varphi$ simply shifts the phase of the signal to $\varphi \rightarrow \varphi+\delta\varphi$ and the oscillations become linear in $\ln k$-space with a clock frequency $(m_\sigma/H)$.
Ekpyrotic scenarios correspond to $0<p<1/3$ and matter contraction to $p=3/2$.

Current Planck data does not show preference for any values of $p$ as no clock signal is detected \cite{Chen:2014cwa,Hamann:2021eyw}.
It is therefore crucial to investigate whether we will be able to do that and to rule out alternative scenarios to inflation based on measurements of the clock signal. 
The procedure we adopt is the following. We take our CPSC best fit. From our forecast analysis
above, we know that we will be able to find very strong to decisive evidence in favor of this model if this is the true model.
By looking at the reconstructed power spectrum and at the CMB residuals, we can identify the beginning of the clock as $\ell_r\sim700$.
Therefore, we fit the template in Eq.~\eqref{eq:clock_template} to the fiducial considering only $\ell>700$. The goal is to check whether we can get a constraint on the parameter $p$.
To make the analysis faster, and to avoid degeneracies with the cosmological parameters that will be induced by cutting $\ell<700$, only in this analysis we fix $\omega_c,\,\omega_b,\,100\theta_s,\,\tau$ to the CPSC best fit values, but we do vary the amplitude and tilt of the power spectrum. For simplicity we only explore the most optimistic experimental configuration, i.e. PL + LB + S4.
We use the following priors on the clock parameters (the priors on the amplitude, tilt and frequency $m_\sigma/H$ are left unchanged): $\log_{10} \, k_r\in [-1.25,\,0],\,$ $\varphi \in[0,\,2\pi],\,$ $A\in[0,\,0.3],\,$  $\log_{10} ( \lvert p\rvert)\in[-3,\,2.7]$. 
We perform two runs, one for positive and one for negative values of $p$. Since inflation corresponds to large values of $\lvert p\rvert\gg1$ and we want to exclude values $\lvert p \rvert<1$, it is very useful to adopt a logarithmic prior on $p$. However, it is complicated to do so for a parameter that switches sign. 

The results are shown in Fig.~\ref{fig:alternative_scenarios}, where we only plot the posteriors for the relevant parameters, i.e.~the frequency and amplitude of the signal together with $p$. As can be seen, we will get a lower bound on the parameter $p$, i.e.~$\log_{10}\, p  > 0.328$ and $\log_{10}\, (-p)  > 0.323$ at 3$\sigma$ for positive and negative values of $p$ respectively.
As also shown in Fig.~\ref{fig:alternative_scenarios}, this will rule out alternative scenarios such as Ekpyrosis or matter contraction in a model independent way which is complementary to the detection of B-modes, traditionally considered the smoking gun of Inflation.

We also comment that the above method is suggestive but incomplete, because a complete method of comparing CPSC signals of different scenarios should use full predictions that include both sharp feature and clock signals. However, for alternative scenarios to inflation, these results are unavailable to date.

			\section{Very similar features: the case for primordial non-Gaussianities}
            \label{sec:comparison2}
			\begin{figure}
\begin{center}	\includegraphics[width=.45\columnwidth]{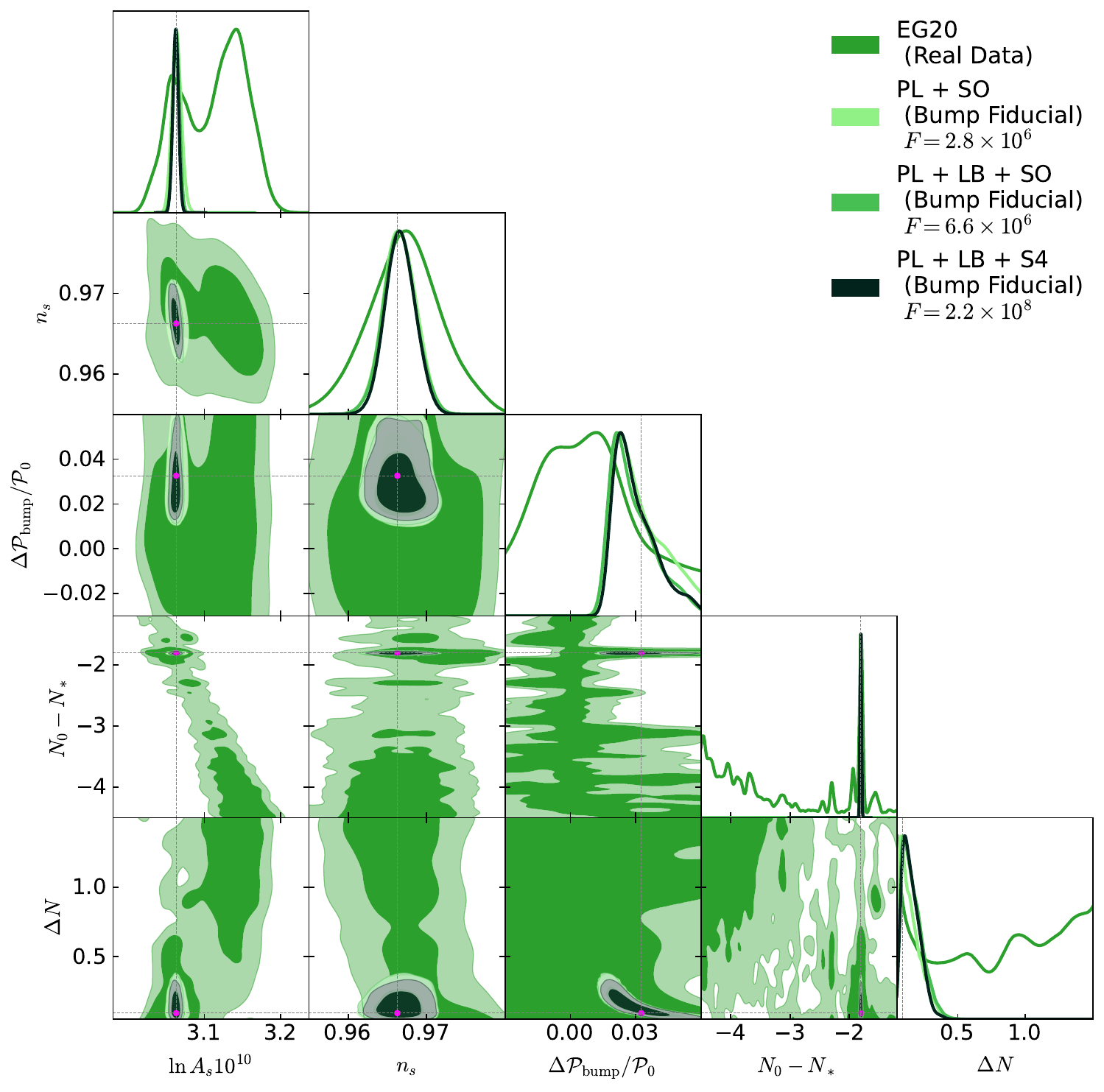}
	\includegraphics[width=.45\columnwidth]{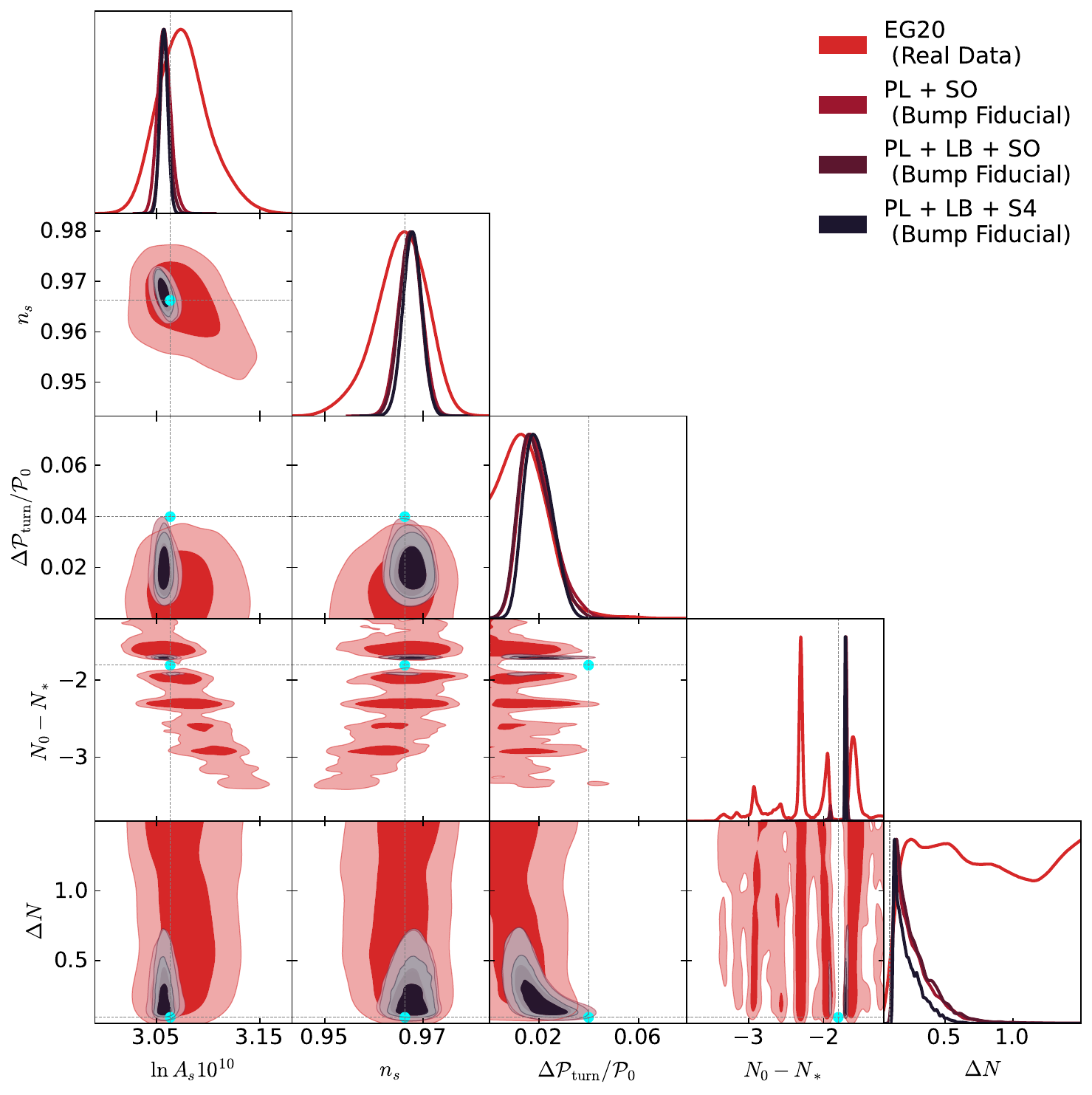}
	\includegraphics[width=.45\columnwidth]{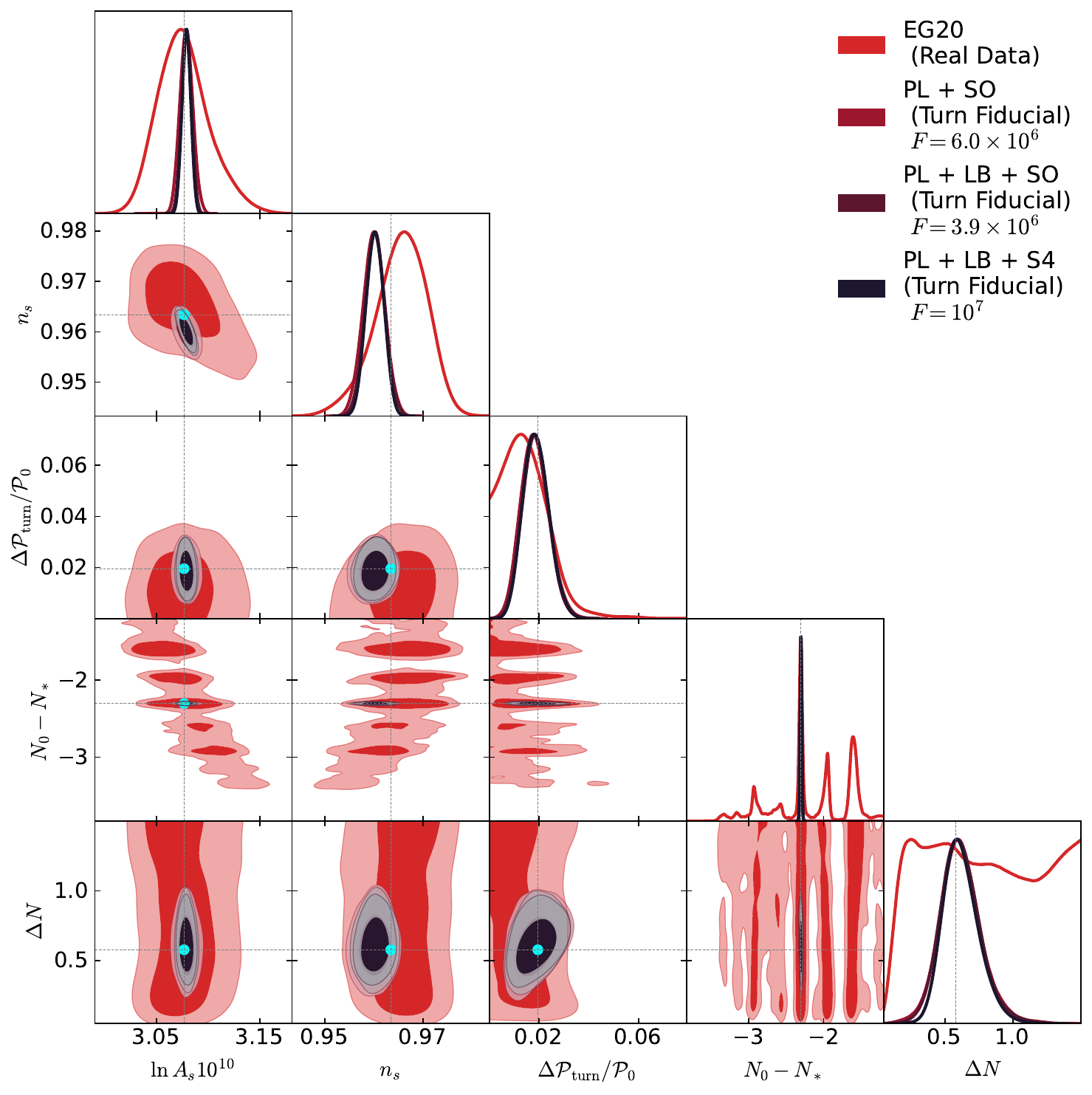}
	\includegraphics[width=.45\columnwidth]{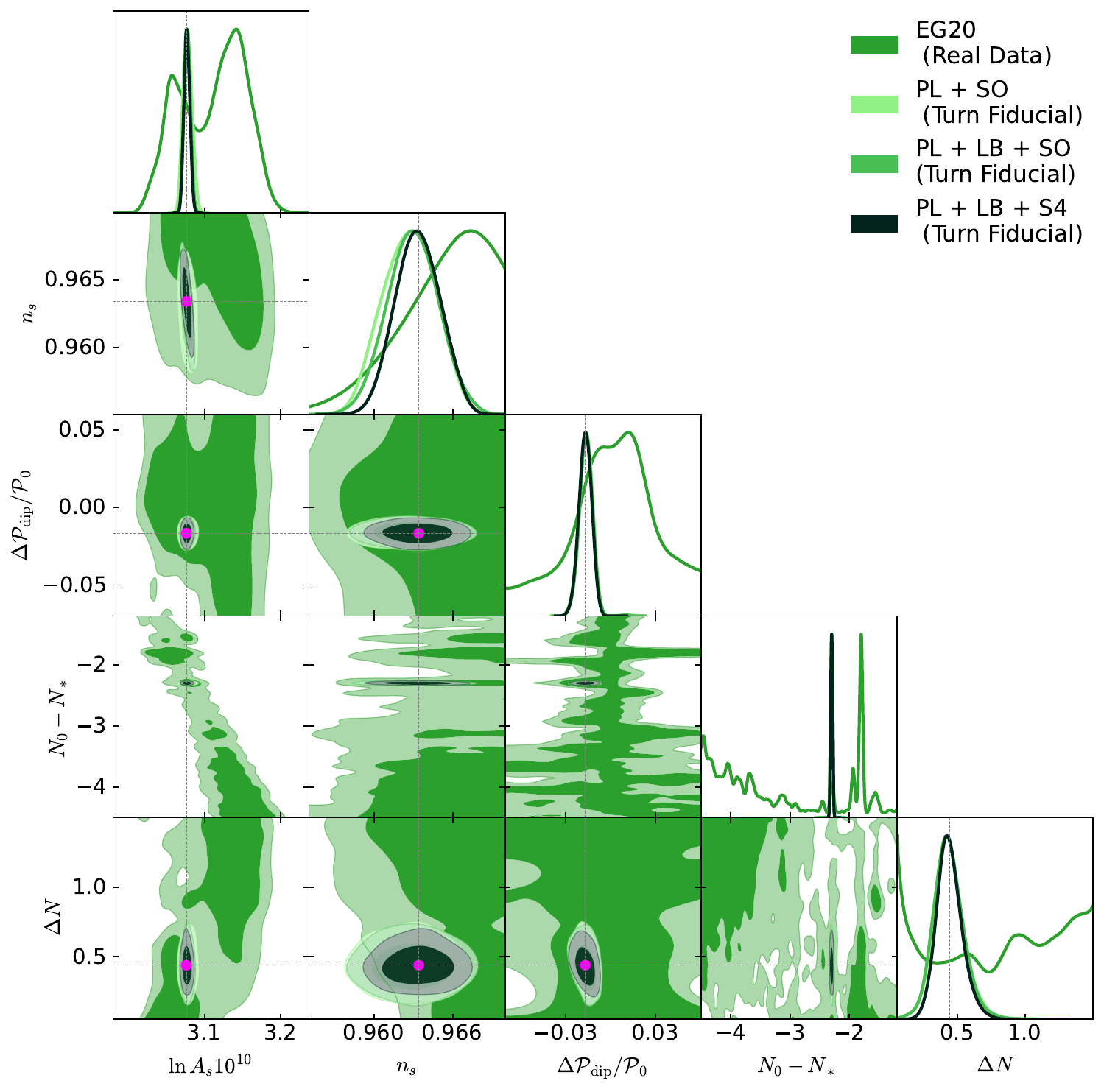}
	\includegraphics[width=.9\columnwidth]{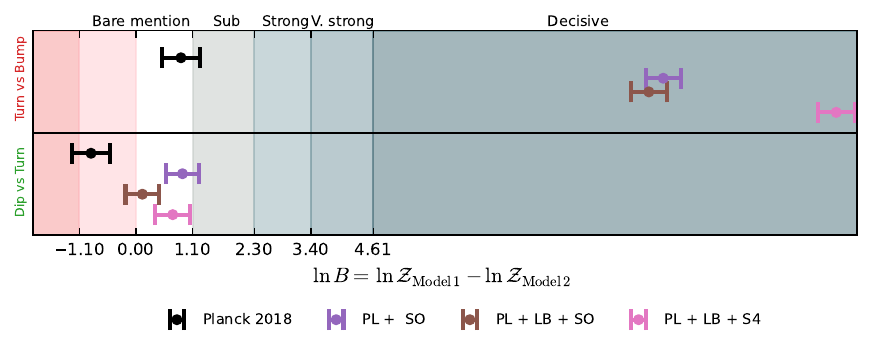}
	\end{center}
	\caption{\footnotesize\label{fig:bump_forecast} 
	Constraints on the Bump and Turn models assuming the bump fiducial (top) and for the Turn and Dip models assuming the turn fiducial (bottom). We also report the Bayes factors from our analysis in the bottom panel. 
	}
	
\end{figure}

In the previous Section, using the fiducial CPSC model as a representative example, we saw that the prospects for distinguishing features of different nature are promising. 
As a final example, we address the following question: can we distinguish, based only on the power spectra, between two feature models that can produce very similar features at high $\ell$?

To answer that, we focus on the Bump/Dip and the Turn models. We remind the reader that a dip in the single-field potential produces a sharp feature signal which is very similar to that produced by a sharp turn in the field space of a multifield model of inflation, except at the largest scales where nevertheless its amplitude is small.
It is therefore interesting to investigate if the  precision of future experiments is enough to distinguish between those two models based only on such a small difference at large scales.
We also remind that the Bump model produces features with opposite phase compared to the Dip and the Turn models.

There are therefore two interesting situations. First, if the Bump best fit is the true model of our Universe, we should be able to rule out the Turn model.\footnote{
Here we take Bump instead of Dip as the fiducial because, if we fit Bump/Dip model to data, currently the data prefers Bump.
}
Second, if the Turn best fit is the true model of our Universe, we expect the Dip model to fit it quite well (again, except for the largest scales).
In the following, we will refer to these two cases as TB (Turn vs Bump) and DT (Dip vs Turn) respectively. The results for TB and DT are shown in the top and bottom panels of Fig.~\ref{fig:bump_forecast} respectively.
In the same figure, we also provide the Bayes factors from our analyses.

\paragraph{Turn v.s. Bump (TB).} Let us first discuss the TB case.
The Bump model can be constrained very well, and we see a detection of all the parameters.
Interestingly, because a feature is detected with very high significance, the parameters of the Turn model are very well constrained, even though it cannot fit the data as well as the Bump model and is eventually decisively excluded.
The feature parameters of the Turn model are constrained to (we take PL + LB + S4 constraints for illustration) $N_0-N* = -1.712\pm 0.012 ,\,$ $
\Delta N_{\rm turn} = 0.228^{+0.031}_{-0.11}\simeq\,\Delta N_{\rm bump}\,\sqrt{2} ^{+0.031}_{-0.11},\,$ $
\Delta \mathcal{P}_{\rm turn}/\mathcal{P}_0 = 0.0197^{+0.0048}_{-0.0064}$.
We thus have a detection of non-zero feature amplitude, although the mean value is smaller than the bump one:   $ 	\Delta \mathcal{P}_{\rm bump}/\mathcal{P}_0 = 0.0325^{+0.0066}_{-0.016}$.
Note that, apart for the difference in the phase of the oscillations that we already mentioned, another fact explains why the sharp feature from a turn is unable to fit the one from the Bump best fit.
The Bump best fit is described by a very sharp feature, with $\Delta N=0.09$.
For such a small value of $\Delta N$, needed to match the envelope of the sinusoidal oscillations, the Turn model develops a quite large bump at large scales, spoiling even more the fit to the fiducial.
Despite the hint to a best fit, the $\chi^2$ to the data is much worse for the Turn model than for the Bump one.
This, together with a large fraction of the prior volume which ends up wasted, leads to a very bad Bayes factor for the Turn model, and therefore the Bump model is decisively favored.
We thus provided one with an example where two models, which are degenerate in a fraction of 
parameter space, can be told apart with a very high statistical significance.
Physically, our results imply that we would be able to reconstruct
a tiny bump in a single-field inflationary potential, without the risk to confuse it with a sharp turn in a multi-dimensional field space.

\paragraph{Dip v.s. Turn (DT).} We now turn to the second case, i.e DT.
The situation is now very different since the Dip model is able to fit virtually all the feature signal produced by the Turn one in the primordial power spectrum, see the left panel of Fig.~\ref{fig:NG}.
More in detail, as can be seen in the lower panel of Fig.~\ref{fig:bump_forecast}, both amplitudes are constrained to a very similar value $\Delta \mathcal{ P}_{\rm turn}/\mathcal{P}_0 = 0.0197^{+0.0048}_{-0.0064}$ and  $\Delta\mathcal{ P}_{\rm bump}/\mathcal{P}_0 = -0.0167\pm 0.0042 $, where again we quote  results from PL + LB + S4.
The mean values of $N_0$ and $\Delta N$ also match, keeping in mind that to get the same feature envelope we must have $\Delta N_{\rm turn} = \Delta N_{\rm bump}\,\sqrt{2}$. The two models will thus be indistinguishable at the level of the power spectrum, as can also be seen from the Bayes factors in Fig.~\ref{fig:bump_forecast}.
Physically, our results imply that, \textit{using only the power spectrum}, we would not be able to claim the detection of a sharp turn in a multi-dimensional field space, as the same feature could be produced by a tiny dip in a single-field inflationary potential.

\begin{figure}
	\includegraphics[width=.5\columnwidth]{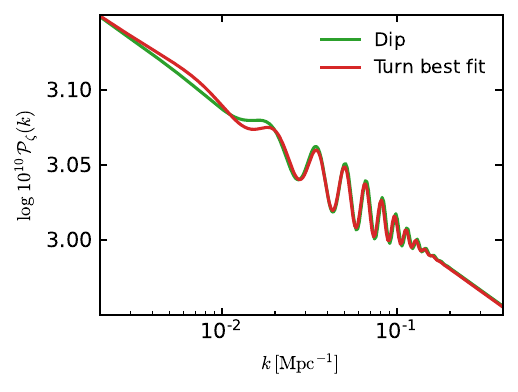}
	\includegraphics[width=.49\columnwidth]{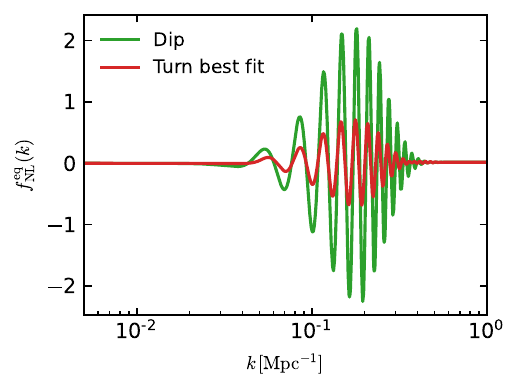}
	\caption{\footnotesize\label{fig:NG} [Left] Power spectra for the best fit Turn and for a Dip model matching the oscillations after the first one. They are the best fit from the TD analysis. [Right] Associated $f_{\rm NL}^\mathrm{eq}(k)$ in the equilateral limit. 
	}
\end{figure}

\paragraph{Primordial non-Gaussianities.} Of course, it would be desirable to distinguish the Turn model from the Dip one in order to draw conclusions about the micro-physical details of inflation.
Since we have just shown that this was impossible using only the power spectrum, we turn to the comparison of the oscillatory features these models leave in the primordial three-point function, the bispectrum (see, e.g., Refs.~\cite{Chen:2006xjb,Chen:2008wn}).
Using the transport approach implemented in the {\tt PyTransport} code~\cite{Dias:2016rjq,Mulryne:2016mzv}, we numerically compute the value of the bispectrum for the two best fits.
More precisely, the bispectrum $B_\zeta(k_1,k_2,k_3)$ is defined as $\braket{\hat{\zeta}_{\vec{k}_1}\hat{\zeta}_{\vec{k}_2}\hat{\zeta}_{\vec{k}_3}} = (2\pi)^3 \delta^{(3)}\left(\vec{k}_1+\vec{k}_2+\vec{k}_3\right) B_\zeta(k_1,k_2,k_3)$, and we vary the overall scale $k$ of the equilateral configuration corresponding to $k_1=k_2=k_3=k$, quoting the dimensionless number $f_\mathrm{NL}^\mathrm{eq}(k)=(10/9) \times k^6 B_\zeta(k,k,k) /\left[(2\pi)^4\mathcal{P}_0(k)^2\right]$.
Note that in this definition we have used the power-law power spectrum $\mathcal{P}_0(k)$ containing the $n_s$ scale-dependence but not the feature one.
The scale-dependent quantities $f_\mathrm{NL}^\mathrm{eq}(k)$ for the Turn and the Dip best fits are represented in the right panel of Fig.~\ref{fig:NG}. 
Two remarks are in order.
First, the running of the signal in the equilateral configuration of the bispectrum is very similar in both cases, just like in the power spectrum.
Second, the Dip best fit results in a bispectrum with an amplitude roughly 3 times larger than the Turn one.
Although the $f_\mathrm{NL}^\mathrm{eq}$ of order unity found for the Turn and Dip best fits will not be constrained with upcoming CMB experiments,  these results show that primordial non-Gaussianities may play a crucial role in disentangling models of inflation that produce very similar features in the power spectrum, if they are detectable by observations with higher sensitivities.
We leave for future works a more thorough analysis of primordial non-Gaussianities in the context of the feature models and the comparison methodology presented in this work, in order to correctly conclude about the discriminating power of oscillatory features in higher-order correlation functions in the CMB~\cite{Fergusson:2008ra,Fergusson:2014hya,Fergusson:2014tza,Planck:2019kim}.
In particular, the information contained in the full shape of the bispectrum should be relevant, beyond the exact equilateral configuration.
In this respect, upcoming Large-Scale-Structure experiments like DESI~\cite{DESI:2016fyo}, Euclid~\cite{Amendola:2016saw} and SPHEREx~\cite{Dore:2014cca}, as well as more futuristic hydrogen atom 21cm map experiments may bring a crucial complementary constraining power.

\section{An estimator for correlation between feature models}
\label{sec:estimator}

			\begin{figure}
			
				\begin{center}
					\includegraphics[width=.495\columnwidth]{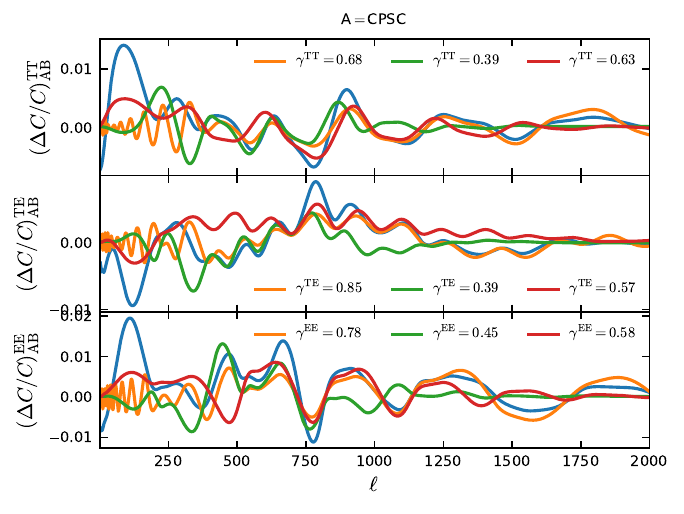}
					\includegraphics[width=.495\columnwidth]{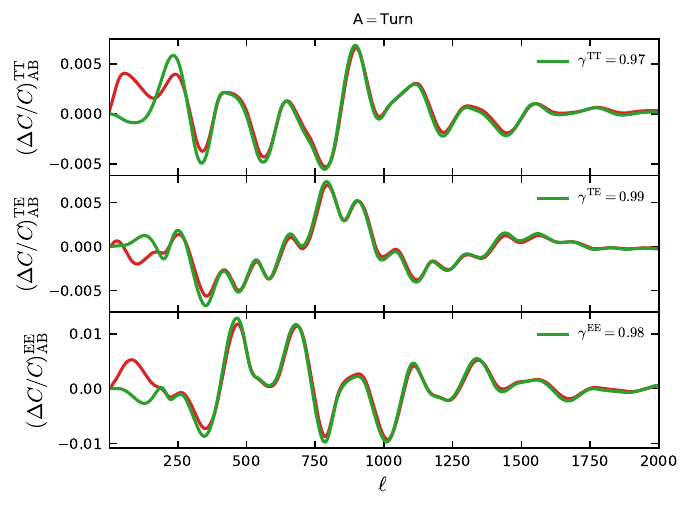}
					\end{center}
								
				\begin{center}
			\footnotesize
			\begin{tabular}{|l|l|l|}
					 \multicolumn{3}{c}{ A = CPSC   $(4.5\,\sigma)$}\\\hline
					              &Forecast (as in Fig.~\ref{fig:cpsc_comparison}) & $\gamma_{A,B}\times$ 4.5$\,\sigma$\\         \hline
					Resonant       & Between 2 and 3$\sigma$ & 3.2$\sigma$ \\         \hline
					Bump/Dip       &  Between 1 and 2$\sigma$& 1.8$\sigma$ \\         \hline
					Turn      &  Between 1 and 2$\sigma$ & 2.7$\sigma$ \\         \hline
		\end{tabular}
		\begin{tabular}{|l|l|l|}
				
					 \multicolumn{3}{c}{ A = Turn   $(3.9\,\sigma)$}\\\hline
				              &Forecast (as in Fig.~\ref{fig:bump_forecast}) & $\gamma_{A,B} \times$3.9$\,\sigma$ \\         \hline
				Dip       & 3.9$\sigma$& 3.9$\,\sigma$ \\         \hline		\end{tabular}
				\end{center}
			
				\caption{\label{fig:estimator} We plot the normalized $\Delta C_\ell^{XY}$. [Top-Left] we take A to be the CPSC bestfit and B as the various best fits to the CPSC fiducial from our forecast. [Top-Right] we take A to be the turn bestfit and B as the bestfit of the bump model to the turn fiducial from our forecast.   The colors are the same as in Fig.~\ref{fig:bestfit_eg20}, i.e.,~blue--CPSC, yellow--resonant, green--bump, red--turn. In the lower tables, using the method of this section, we estimate the detection levels of these models B based on the detection level of the fiducial model A, and compare them to the exact detection level from our forecast for PL+LB+S4. Since the $\gamma$s in the top panels are very similar to each others, we simply estimate the detection levels by using the lowest $\gamma$ for each model.
				}
			\end{figure}

As we have seen from previous examples, some feature models share more similarities than others. It may be useful to construct a simple estimator quantifying how closely two models are correlated.
If a new feature model is constructed that is not analyzed above, one can use this estimator to give a quick estimate of a lower bound of statistical significance based on its correlation with the models analyzed in this work, without having to perform the full data analysis.

We define a correlator $\gamma$ between two different feature profiles, $\Delta \mathcal{P}_A$ and $\Delta \mathcal{P}_B$, as
\begin{equation}
	\gamma^{XY}_{A,B} \equiv
	\frac{(\Delta \mathcal{P}_A, \Delta \mathcal{P}_B)}
	{\sqrt{(\Delta \mathcal{P}_A, \Delta \mathcal{P}_A)(\Delta \mathcal{P}_B, \Delta \mathcal{P}_B)}} ~,
	\label{Eq:correlator_def}
\end{equation}
where the scalar product $(\cdot\,,\,\cdot)$ is defined as
\begin{equation}
    (\Delta \mathcal{P}_A, \Delta \mathcal{P}_B) \equiv
    \int_{\ell_{\rm min}}^{\ell_{\rm max}}
    \mathrm{d} \ell ~ w(\ell) ~ 
    \frac{\Delta C^{XY}_{\ell, A}}{C_\ell^{XY}}
    \frac{\Delta C^{XY}_{\ell, B}}{C_\ell^{XY}} ~,
\end{equation}
in which $XY$ stands for $TT$, $TE$ or $EE$.
$\Delta C_\ell^{XY}$ is computed through the transfer function $T^{XY}_\ell(k)$,
\begin{equation}
    \Delta C_\ell^{XY} = \int \mathrm{d} \log k ~ T^{XY}_\ell(k) \Delta\mathcal{P} ~.
\end{equation}
Also note that, in the case $X,\,Y= T,\, E$, instead of considering $C_\ell^{TE}$ at the denominator which vanishes at some multipoles, we use the quantity $\sqrt{C_\ell^{TE}\,C_\ell^{TE} + C_\ell^{TT}\,C_\ell^{EE}}$. The weight function $w(\ell)$ is determined by the sensitivity of the experiment. Since both the temperature and polarization will be eventually measured up to cosmic variance limited precision, we will assume the cosmic variance limit when defining the weight function. As the signal-to-noise ratio of the fractional correction to the CMB power spectrum introduced by each feature is given by $\left( \Delta C_{\ell}^{XY}/C_\ell^{XY} \right)/\left( 1/\sqrt{2\ell+1} \right)$, the weight function can be chosen as
\begin{equation}
w(\ell)= 2\ell+1~.     
\end{equation}

Several examples of this correlation estimator are shown in the top panels of Fig.~\ref{fig:estimator}, where we also plot the quantities $\Delta C^{XY}_{\ell, A, B}/C^{XY}_{\ell, A, B}$ for a better understanding of the results. Combined with Fig.~\ref{fig:cpsc_comparison} and \ref{fig:bump_forecast}, from these examples we see that, in the event of a feature detection, we will be able to distinguish two models using CMB even if they are up to $\sim 80\%$ correlated, although we caution that there may be model-dependent exceptions.

Another use of this estimator is the following. If a model A can be detected in the CMB XY data with a statistical significance of $x^{XY}$-sigma and a new model B, which has not been compared with data, is $\gamma_{A,B}^{XY}$ correlated with A, then one can estimate a rough lower bound of 
$|\gamma^{XY}_{A,B}|~x^{XY}$-sigma detection for model B with the XY data. This estimate is for the individual TT, TE or EE data. In the analyses of previous sections, the constraints we got combined TT, TE and EE, so, without knowing the individual constraints on model A, generally speaking, we do not know how to get the combined estimate for model B. On the other hand, in a simple situation where all three $\gamma_{A,B}^{XY}$'s are similar $\sim \gamma_{A,B}$ (as in the several examples shown in Fig.~\ref{fig:estimator}), we can approximate the lower bound as $|\gamma_{A,B}|~x$-sigma if the combined significance for model A is $x$-sigma. Since we already have done the model comparison in the CPSC and Turn model fiducial data, in the tables in Fig.~\ref{fig:estimator} we test the values of $|\gamma_{A,B}|~x$-sigma for several models.

We have chosen to construct the estimator in terms of the CMB multipole space $\ell$ instead of the momentum space $k$ of the original primordial feature profile, because the signals in the CMB multipole space is what is relevant to measurements. This choice is experiment specific; if we are concerned with galaxy survey experiments instead of CMB, the scalar product can be defined in the 3D momentum space,
\begin{equation}
	(\Delta \mathcal{P}_A, \Delta \mathcal{P}_B) =\int_{k_{\rm min}}^{k_{\rm max}}\mathrm{d}k~w_{\rm LSS}(k) \frac{\Delta P_A}{P}(k)\,\frac{\Delta P_B}{P}(k),
\end{equation}
with a different weight function $w_{\rm LSS}$.

			\section{Conclusions}
			\label{sec:conclusions}
			
Primordial features carry rich information on the primordial universe and can potentially deliver answers to some of the most important questions about inflation and alternative theories.
These properties have been pointed out, studied and classified in many recent works \cite{Chen:2010xka,Chluba:2015bqa,Slosar:2019gvt,Achucarro:2022qrl}. Unfortunately, theoretically, the amplitudes of such signals have no generic lower bound and are highly model-dependent, and experimentally no statistically significant signals have been detected.
Thus, the future of this field critically depends on the prospects of observational sensitivities for these signals in the near future.
In this paper, we addressed this question in the context of CMB E-mode polarization data, which will soon be updated by experiments such as Simons Observatory, LiteBIRD, and CMB S4.
Due to their sharper transfer functions, E-modes carry more information about features than T-modes and are therefore expected to improve our knowledge about features.

In this work, we first summarized the theoretical predictions of several representative feature models on the power spectrum using a unified formalism.
All these models currently give good fits to some small-scale glitches in the Planck residual data.
Then, using a state-of-the-art pipeline optimized to compare feature models with data and using Bayesian inference, we showed that, if the featureless model is the true model of the universe, these feature candidates from the Planck data will be ruled out with strong evidences.
Conversely, if one of these feature models is the true model of the Universe, the featureless one will be ruled out with very strong evidence. 

Next, we addressed the main question of this paper.
If one of the candidate feature models is confirmed by upcoming CMB polarization data, we will have to consider the possibility that this feature can be explained by several models that have very different underlying physics.
Will these data be strong enough to tell the differences?
We found that, due to the rich information that feature-model predictions carry with their phases and envelops across all scales, future observations will be able to distinguish them with very strong to decisive Bayesian evidences.
This result is even more impressive when we realize that the feature models we considered are often highly correlated to each others.
However, there do exist exceptional cases in which microscopically very different models predict almost identical features in the power spectrum.
Still, we showed for an example, that two models with identical power spectra lead to primordial non-Gaussianities of different amplitudes.
A more thorough analysis of the bispectrum shape functions for these very similar feature models, and the study of the detectability of these non-Gaussian signals, are left for future study.

These findings
clearly demonstrate that the study of primordial features should be considered
as one of the main 
tools to understand the primordial universe in the next decades.
With the addition of the large-scale structure data from various galaxy surveys~\cite{Huang:2012mr,Hazra:2012vs,Chen:2016vvw,Ballardini:2016hpi,Palma:2017wxu,LHuillier:2017lgm,Ballardini:2017qwq,Beutler:2019ojk,Ballardini:2019tuc,Debono:2020emh,Li:2021jvz,Ballardini:2022wzu,Chandra:2022utq} in a similar time frame, as well as 21cm hydrogen line surveys~\cite{Chen:2016zuu,Xu:2016kwz} and stochastic gravitational wave background mapping \cite{Fumagalli:2020nvq,Braglia:2020taf,Bodas:2022zca} in a more distant future, we can hope for discoveries that will shake our understanding of the origin of the Universe.

\medskip
\section*{Acknowledgments}
			
MB is supported by the Spanish Atracci\'on de Talento contract no. 2019-T1/TIC-13177 granted by Comunidad de Madrid, the I+D grant PID2020-118159GA-C42 of the Spanish Ministry of Science and Innovation and the i-LINK 2021 grant LINKA20416 of CSIC. MB and DH thanks the Indian Institute of Technology Madras, where part of this work was carried out, for hospitality and acknowledge travel support through the India-Italy
mobility program (INT/Italy/P-39/2022 (ER)).
L.P. would like to acknowledge support from the “Atracci\'{o}n de Talento” grant 2019-T1/TIC15784, his work is partially supported by the Spanish Research Agency (Agencia Estatal de Investigaci\'{o}n) through the Grant IFT Centro de Excelencia Severo Ochoa No CEX2020-001007-S, funded by MCIN/AEI/10.13039/501100011033.
			
			\appendix
			
	\section{Methodology}
\label{app:analysis}

To solve for the inflationary dynamics, we use BINGO and its two-field extension.
The details of the numerical methods are outlined for single-field models in~\cite{Hazra:2012yn} and for two-field models in~\cite{Braglia:2020fms}.
Note that while for single-field models we integrate the perturbation equations from sub-Hubble to super-Hubble scales, for two-field models we integrate till the $e$-fold when the curvature perturbation freezes, to avoid any inaccuracies due to unaccounted super-Hubble evolution. CAMB~\cite{Lewis:1999bs,Howlett:2012mh} is used to calculate the angular power spectrum.
The primordial power spectrum, for single-field models, is evaluated for all the modes required by CAMB without interpolation, while for two-field models, we use an adaptive interpolation as outlined in~\cite{Braglia:2021ckn}.
We would like to note that we evaluate angular power spectra at all multipoles as our priors in the potential parameters allow for high-frequency oscillations that can be missed if an interpolation is used in the multipole space.

For comparison with data and forecast we use {\tt CosmoChord} ({\tt PolyChord} add-on of {\tt CosmoMC}\footnote{\href{https://cosmologist.info/cosmomc/}{https://cosmologist.info/cosmomc/}} \cite{Lewis:2002ah})~\cite{Handley:2015fda,Handley:2015vkr}. We used the best fits obtained against the~{\tt 2020 CamSpec release v12.5}\footnote{\href{https://people.ast.cam.ac.uk/~stg20/camspec/index.html}{https://people.ast.cam.ac.uk/~stg20/camspec}} (hereafter EG20, from the names of its authors~\cite{Efstathiou:2019mdh}) datasets as this unbinned dataset covers larger sky fraction in temperature and polarization with a different cleaning.
In all analyses, the nuisance parameters are allowed to vary and priors on the nuisance parameters are used accordingly.  The priors on the model parameters are shown in Table~\ref{tab:priors}.

The forecast pipeline follows the methodology discussed in our earlier analysis in~\cite{Braglia:2021rej}. For completeness, we briefly discuss the pipeline here as well. We work with the sensitivities of the upcoming and proposed CMB observations.
The multipole range for the fiducial spectra for the forecasts are fixed to scales probed by Planck, {\it i.e.} between $\ell=2-2500$. Since the best fits to the present data are obtained by comparing the models in this multipole range, our choice ensures conservative forecasts and a comparison of models within the same scales.
Depending on the sensitivities, we combine the present and upcoming observations spanning the multipole range. We consider three combinations PL + SO, PL + LB +SO and PL + LB + S4. For the PL + SO combination, we assume 80\% sky coverage from Planck at large scales ($\ell=2-39$) and 20\% at scales corresponding to $\ell=40-2500$. While Planck already has data with significantly larger sky coverage at small scales, for forecast, in order to avoid the overlap with SO we use a smaller fraction. We assume SO to cover $\ell=40-2500$ with 40\% sky fraction. In PL + LB + SO, keeping the multipole range and sky coverage same for SO, we redistribute the scales between Planck and LiteBIRD. PL noise is used here between $\ell=800-2500$ (20\% sky coverage) and LB noise is used for multipoles $\ell=2-39$ (80\% sky coverage) and $40-1350$ (20\% sky coverage).
For PL + LB + S4 we use the same distributions in the multipole range with the SO noise and sky coverage replaced by that of CMB-S4.
Where the signal-to-noise ratios of these observations are comparable, this distribution ensures insignificant overlap between them.

For all our forecast analyses, an inverted Wishart likelihood has been used following~\cite{Hamimeche:2008ai}. The noise power spectrum is calculated using the r.m.s. noise per pixel, full width at half maxima of the assumed Gaussian beam and the sky fraction. For SO, the baseline noise configuration of the large aperture telescope has been used\footnote{The temperature and polarization noise power spectra for SO that we use in our analysis are available at \href{https://github.com/simonsobs}{https://github.com/simonsobs}.}. Note that we have not used the B-mode noise as we are not considering tensor modes.

 \begin{table}
 	\centering 
 	\begin{tabular}{|l|l|}
 		\hline
    Cosmological 		Parameters               & Priors      \\ \hline
 		$\Omega_\mathrm{b}h^2$   & $[0.02,\, 0.0265]$ \\ \hline
 		$\Omega_\mathrm{CDM}h^2$ & $[0.1,\, 0.135]$   \\ \hline
 		$100*\theta_s$              & $[1.03,\, 1.05]$   \\ \hline
 		$\tau$                    & $[0.03,\, 0.08]$      \\ \hline
 	\end{tabular}
 	 	\begin{tabular}{|l|l|l|l|l|}
 	\hline Feature parameters & \multicolumn{4}{c|}{ Priors}\\	\hline & Resonant & CPSC & Turn  &Bump/Dip\\\hline 		$\ln\,10^{10} A_s$   &\multicolumn{4}{c|}{$[2.763,    \,4.375]$} \\ \hline 		$n_s$   &\multicolumn{4}{c|}{$[0.920,    \,0.996]$} \\ \hline 		$\Delta \mathcal{P}/\mathcal{P}_0$   &\multicolumn{3}{c|}{$[0,    \,0.35]$}&$[-0.35,\,0.35]$ \\\hline 		$N_0-N_*$  &$-$ &\multicolumn{3}{c|}{$[-4.3,    \,-1.0]$} \\	\hline $\log_{10}\,\omega/H$   &\multicolumn{2}{c|}{$[0,    \,1.9]$}    &\multicolumn{2}{c|}{$-$}\\\hline	 $\Delta N$   &\multicolumn{2}{c|}{$-$}    &\multicolumn{2}{c|}{$[0.05,\,1.5]$}\\\hline	
 	$\varphi$  & $ [0,\,2\pi]$ &\multicolumn{3}{c|}{$-$} \\\hline
\end{tabular}
 	\caption{\label{tab:priors} Priors on the cosmological and feature parameters used in our analysis in each models. The priors on foreground and calibration parameters are kept unchanged from their default values in the Polychord analysis. Note that, to unify the nomenclature, we define $\omega=m_\sigma$ for the CPSC model.}
 \end{table}

 To describe the statistical power of future CMB surveys to constrain the parameter space of the models, we use the so called {\em Figure of Merit} (FoM), defined as:
\begin{equation}
	{\rm FoM} \equiv \lvert{\rm Fi}\rvert^{-1/2},
\end{equation} 
where ${\rm Fi}$ is the covariance matrix obtained from our data analysis. An increase in the {\rm FoM} corresponds to a compression of the parameter space going from a survey to another one, so a better constraining power of the latter. We quote 
\begin{equation}
	F\equiv\frac{{\rm FoM}_{\rm experiment}}{{\rm FoM}_{\rm Planck}} .
\end{equation}

			\section{Computation of effective parameters}
			\label{app:analytical_results}
			In this Appendix, we provide a detailed computation of the  feature signals in the power spectrum for the Bump/Dip and the Turn model using the in-in formalism (see e.g.~\cite{Chen:2010xka} for a review). We also compare the analytic prediction with the numerical spectra to show to what extent they agree. The purpose is to derive the effective parameters used in the analysis in the main text. Since the effective parameters of the resonant and CPSC model have been already derived elsewhere, we restrict our discussion to the sharp feature models mentioned above.
			
			We remind readers that the expressions of the effective parameters are approximate relations that serve as an efficient bridge between observables and raw model parameters. The purpose is to facilitate the search of the parameter space, especially for models with a large number of parameters and with complicated relations between observables and model parameters. All theoretical predictions used in the data analyses and forecasts are numerical results and not approximations. The full methodology is described in Ref.~\cite{Braglia:2021ckn}.

			\subsection{Bump/dip in potential}

			This model is defined by $\Xi = 0,\,  \Delta V/V = A\, \exp\left[-\left(\phi-\phi_0\right)^2/\Delta^2\right]$ (see Sec.~\ref{sec:models}). The signal will therefore depend on the width $\Delta$, location $\phi_0$ and   amplitude $A$ of the perturbation to the potential. The width $\Delta$ can be easily related to the time duration of the sharp feature. Assuming that the feature lasts $\Delta \phi = 6 \Delta$, which is approximately $100 \%$ of the area below the Gaussian perturbation to the potential, we obtain $\Delta N = 6\Delta/\sqrt{2\epsilon_0}$. On the other hand, the relation between $\phi_0$ and $N_0$ is easily computed numerically with the help of a shooting algorithm, see e.g.~Refs.~\cite{Braglia:2021ckn,Braglia:2021rej} for details. In this Section, we derive an approximate expression for the maximum amplitude of the feature, which is the only remaining effective parameter that we need for our data analysis. To do that, we use the in-in formalism. Here, we just sketch the calculation referring the reader to reviews, see for example Ref.~\cite{Chen:2010xka} for the technicalities.
			
			To compute the correction to the SFSR power spectrum, we need to separate the second order Hamiltonian for the curvature perturbation, i.e.
			\begin{equation}
				H_2(\epsilon)=\int\,\mathrm{d}^3 \vec{x}\,a\, \epsilon\, M_{\mathrm{Pl}}^2\left[\zeta'^2+ \left(\partial\zeta\right)^2\right].
			\end{equation}
			into a free and interacting parts defined as
			 $H_{2,\,0}\equiv H_2(\epsilon_0)$ and
			$H_{\text{int},2}\equiv H_2(\Delta\epsilon)$ respectively, where $\epsilon\equiv \epsilon_0+\Delta\epsilon$ and $\Delta\epsilon$ is a small perturbation.
			During the bump/dip, we can ignore the Hubble friction term in equation of motion for the inflaton and we have the simpler energy conservation relation $\Delta (\dot\phi/2 + V)=0$. With the definition $\epsilon=\dot\phi^2/(2H^2)$, we can easily compute $\Delta \epsilon$ as:

						\begin{equation}
				\Delta\epsilon=-3A\exp\left[-\frac{(\phi-\phi_0)^2}{\Delta^2}\right]=-3A\exp\left[-\alpha(x-1)^2\right],
			\end{equation} 
			where $\alpha=\frac{2\epsilon_0}{\Delta^2}=\frac{36}{\Delta N^2}$ and $x\equiv\tau/\tau_0$, where $\tau_0$ is the conformal time of the sharp feature. Using the in-in formalism, we compute the correction to the primordial power spectrum as:
			\begin{equation}
				\label{eq:bump}
				\frac{\Delta \mathcal{P}(k)}{\mathcal{P}_0}=\frac{k^3}{2\pi^2}\frac{1}{\mathcal{P}_0}(-i)\int_{-\infty}^{0_-}{\rm d}\tau a\,\langle0\lvert\left[\zeta(k,\,0)\zeta(k,\,0),\,H_{{\rm int},\,2}\right]\rvert0\rangle,
			\end{equation}
		where the state $\lvert 0\rangle$ is the state annihilated by the annihilation operators in the expansion of the curvature perturbation $\zeta$. Note that $\zeta$ is in the interaction picture where operators are evolved using the free Hamiltonian $H_{2,\,0}$.

			Using the expression for $\Delta\epsilon$ found above, the integral becomes:

			\begin{equation}
				\frac{\Delta \mathcal{P}(k)}{\mathcal{P}_0}=-\frac{3A}{\epsilon_0}{\rm Im}\int^{\infty}_{0_+}{\rm d}x \,\frac{1}{ x^2}e^{-\alpha (x-1)^2}e^ {2 i x x_0}\left(2 x_0 x^2+2 i x-\frac{1}{x x_0}\right)
			\end{equation}
			where $x_0 \equiv \lvert k\tau_0\rvert$. The integral has a saddle point at $x=1$. So we can simply evaluate the slow-varying terms inside the round brackets at $x=1$ and take them out the integral.
			The remaining integral is simply the integral of a Gaussian once we complete the square in the argument of the exponential. 
			The result is just:
			\begin{equation}
				\label{eq:bump2}
				\frac{\Delta \mathcal{P}(x_0)}{\mathcal{P}_0}=-\frac{3 A}{\epsilon_0}\Delta\sqrt{\frac{\pi}{2\epsilon_0}}\left[2\cos 2 x_0+(2 x_0-\frac{1}{x_0})\sin 2 x_0\right]e^{-x_0^2/\alpha}.
			\end{equation}
			\begin{figure}
				\begin{center}
					\includegraphics[width=.327\columnwidth]{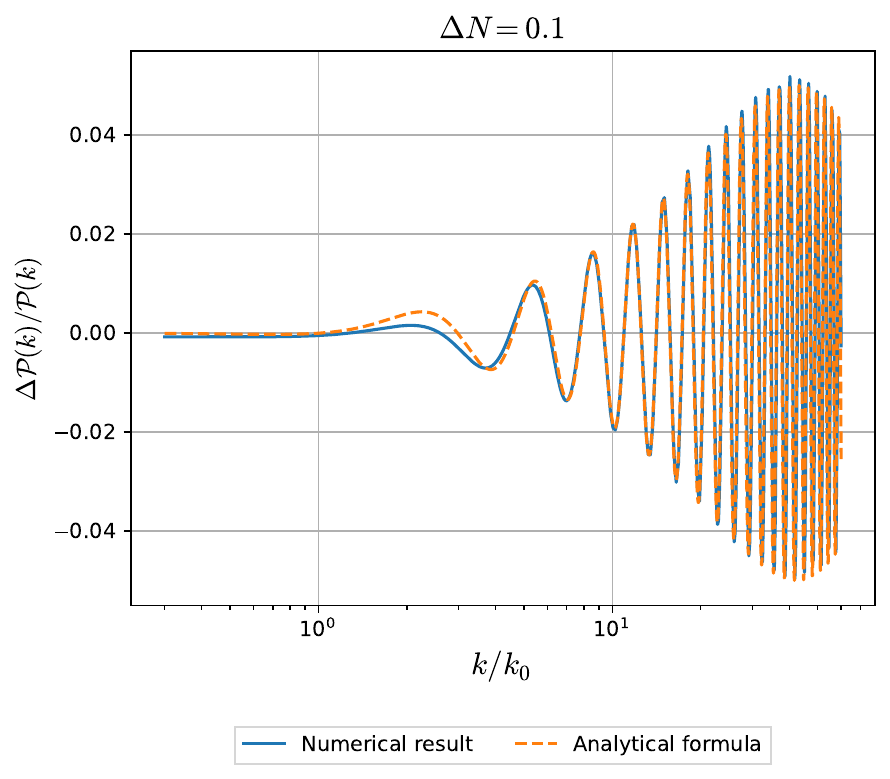}
					\includegraphics[width=.327\columnwidth]{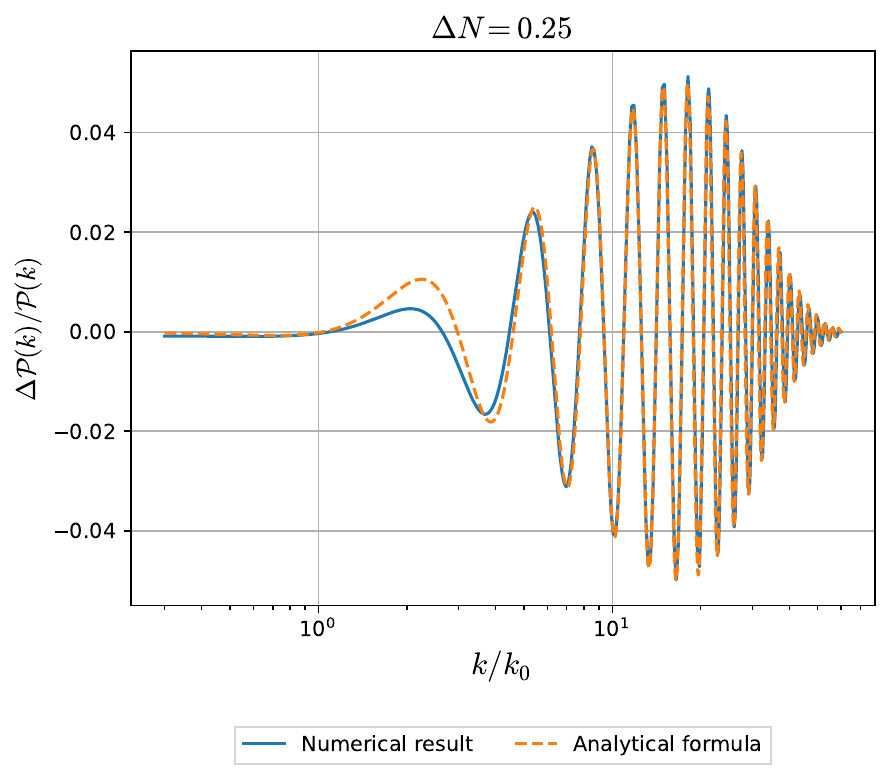}
					\includegraphics[width=.327\columnwidth]{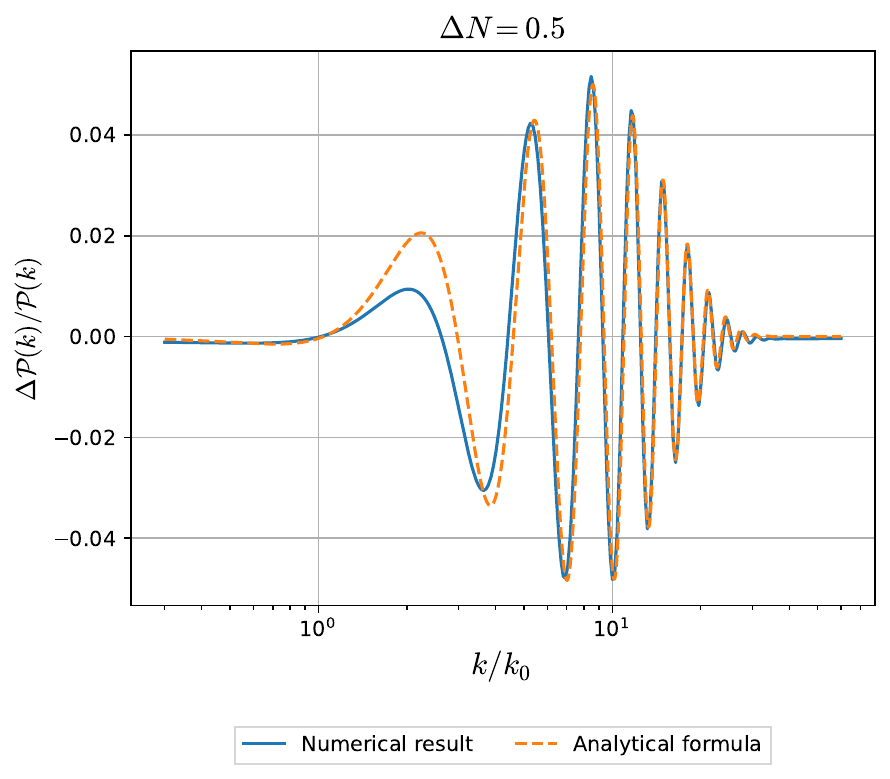}
				\end{center}
				\caption{\label{fig:bump_analytical} Comparison of the analytical formula Eq.~\eqref{eq:bump2}. }
			\end{figure}
			
			Fig.~\ref{fig:bump_analytical} shows the comparison between  the analytical formula and the numerical results. Apart from some a slight discrepancy for $x_0\sim1$, probably due to the saddle point approximation we use,  the two match very well.
			
			Note that the maximum amplitude of the feature depends on $A$ through a simple multiplicative factor. However the dependence on $\alpha$ is more complicated. In fact, the spectrum peaks at different values depending on the value of $\alpha$ through the relation $x_0^{\rm peak}=\sqrt{\alpha/2}$. Evaluating the expression above at $x_0^{\rm peak}$, we get 
			\begin{equation}
				\frac{\Delta P_{\rm max}}{ P_0} = \frac{3 A}{\epsilon_0}\sqrt{\frac{2\pi}{e}}.
			\end{equation}
			This is the effective parameter that we use in our analysis.

			\subsection{Turn model}

				This model is defined by $\Xi = \xi\exp\left[-\left(\phi-\phi_0\right)^2/\Delta^2\right], \, \Delta V/V = 0$ (see Sec.~\ref{sec:models}). Also in this case, the signal will depend on the width $\Delta$, location $\phi_0$ and   amplitude $\xi$ of the perturbation to the kinetic term of the inflaton. As before, the width $\Delta$ can be easily related to the time duration of the sharp feature. Assuming that the feature lasts $\Delta \phi = 6 \Delta$, we obtain again $\Delta N = 6\Delta/\sqrt{2\epsilon_0}$.  To derive an approximate expression for the maximum amplitude of the feature, using the in-in formalism, we integrate out the massive field $\sigma$ and use an effectively single field description, where the second order interacting Hamiltonian now contains a new term, in addition to the one computed previously due to $\Delta\epsilon$:
				\begin{equation}
					H_{{\rm int},\,2}\subset\int\,\mathrm{d}^3 \vec{x}\,a\, \epsilon_0\, M_{\mathrm{Pl}}^2\, \frac{\left(\partial\zeta\right)^2}{c_s^2},
				\end{equation}
where $c_s^2$ is the squared speed of sound of the inflaton, which we compute below as in \cite{Tolley:2009fg, Achucarro:2010da}.

			We start by deriving analytical predictions for the background.  We set the massive field initially at rest at the bottom of its potential $\sigma_i=0$. Eventually, when $\phi\sim\phi_0$, it gets  momentarily displaced from the minimum. Neglecting the acceleration and friction terms in the Klein-Gordon equation for $\sigma$, we get\footnote{We define $f\equiv\left[1+\sigma\Xi(\phi)\right]$.}:
			\begin{equation}
				V_\sigma=m_\sigma^2\sigma\simeq\Xi(\phi)f\dot{\phi}^2=2 H^2 \epsilon_0\Xi(\phi)/f
			\end{equation}
			where we used that $\epsilon_0 = f^2\dot{\phi}^2/2 H^2$.
			A crucial quantity  is the bending rate $\eta_\perp$, wich measures the acceleration perpendicular to the background trajectory.
			Its expression is given by:
			\begin{equation}			H^2\eta_\perp^2=\frac{V_{s}^2}{\dot{\sigma}_A^2}
			\end{equation} 
			where $\dot{\sigma}_A = \sqrt{f^2\dot{\phi}^2+\dot{\sigma}^2}$ and $V_{s}\equiv e_s^I V_I$ where $I=\sigma,\,\phi$ and $e_s^I$ is the entropic vielbein, i.e the unit vector orthogonal to the adiabatic vielbein. In our model, the bending is given by:
			\begin{equation}
				H^2\eta_\perp^2=\frac{1}{\dot{\sigma}_A^4}\left[V_\sigma^2 f^2\dot{\phi}^2-2 V_\phi V_\sigma\dot{\phi}\dot{\sigma}+V_\sigma^2\dot{\sigma}^2f^{-2}\right]\,.
			\end{equation}
			Using the slow-roll approximation for $\phi$, this becomes
			\begin{equation}	H^2\eta_\perp^2=\frac{f^2}{\dot{\phi}^2}\left[V_\sigma-3 H\dot{\sigma}\right]^2.
			\end{equation}
			We can further assume that  $3 H \dot{\sigma}\ll V_\sigma$, to arrive to a very simple expression\footnote{Note that the assumption   $3 H \dot{\sigma}\ll V_\sigma$ is not guaranteed to be always satisfied, but it is so when the turn is not so sharp and it lasts shorter than the typical oscillation period of the massive field.}:
			\begin{equation}
				\label{eq:etaperp}
				H^2\eta_\perp^2\simeq\frac{f^2}{\dot{\phi}^2}V_\sigma^2\simeq 2 H^2 \Xi(\phi)^2\epsilon_0f^2\simeq2 H^2\epsilon_0\xi^2\exp\left[-\frac{2\left(\phi-\phi_0\right)^2}{\Delta^2}\right].
			\end{equation}
			
			We investigate the agreement of this analytical expression with numerical results in the top 3 panels of Fig.~\ref{fig:turn_analytical}. As can be seen, Eq.~\eqref{eq:etaperp} matches quite well the numerics unless $\Delta N$ is very small. The analytical approximation for the feature in the power spectrum that we are going to present in the following does not apply to that case, for which the approximations that we will assume break down. Nevertheless, we do consider spectra that are not captured by our analytical approximation in our data analysis, which is based on numerical integration.
			
			In the single-field effective field theory (EFT),  the turn in the trajectory induces a sudden drop in the speed of sound of the inflaton. Integrating out the  entropic perturbation, well approximated by the $\sigma$ direction in field space, $\zeta$ acquires a speed of sound given by:
			\begin{equation}
				\frac{1}{c_s^2}-1=4\frac{H^2\eta_\perp^2}{m_s^2},
			\end{equation}
			where 
			\begin{equation}
				m_s^2=V_{ss}-H^2\eta_\perp^2+\epsilon H^2 M_{\rm Pl}^2 R_{\rm fs}
			\end{equation}
			is the squared entropic mass and $R_{\rm fs}$ is the curvature of the field-space metric, which is zero in our case $R_{\rm fs}=0$. Note that the last term, which represents the curvature of the field space identically vanishes in our model. The single-field effective theory is valid as long as $m_s^2/H^2\gg1/c_s^2$ and $\dot{c_s}/c_s\ll m_s$~(see e.g. Refs.~\cite{Tolley:2009fg,Cespedes:2012hu,Garcia-Saenz:2019njm,Pinol:2020kvw}). These assumptions are very well satisfied in the center and right panels of Fig.~\ref{fig:turn_analytical}, but we have not checked their validity for the left panel. In fact, in that case, the analytical expression for $H\eta_\perp$ is drastically different from the numerical one, so, even if the EFT held, we would still expect the analytical result obtained with \eqref{eq:etaperp} not to describe the numerical one.
			The explicit expression for the speed of sound in our model is:
			\begin{align}
				\frac{1}{c_s^2}-1=4\frac{H^2\eta_\perp^2}{m_s^2}&\simeq 8 \left(\frac{\xi}{m_\sigma/H}\right)^2\epsilon_0\exp\left[-\frac{2\left(\phi-\phi_0\right)^2}{\Delta^2}\right] \notag\\
				&\simeq8 \left(\frac{\xi}{m_\sigma/H}\right)^2\epsilon_0\exp\left[-\frac{4\epsilon_0\left(x-1\right)^2}{\Delta^2}\right]. 
			\end{align}

			The correction to $\epsilon$ can be computed in a similar way and we get 
			\begin{equation}
				\left(\frac{1}{c_s^2}-1\right)=-4\frac{\Delta\epsilon}{\epsilon_0}
			\end{equation}
			The correction to the power spectrum is
			\begin{equation}
				\label{eq:turn}
				\frac{\Delta \mathcal{P}(k)}{\mathcal{P}_0}=\frac{k^3}{2\pi^2}\frac{1}{\mathcal{P}_0}(-i)\int_{-\infty}^{0_-}{\rm d}\tau a\,\langle0\lvert\left[\zeta(k,\,0)\zeta(k,\,0),\,H_{{\rm 2},\,\Delta\epsilon}+H_{{\rm 2},\,\Delta c_s}\right]\rvert0\rangle,
			\end{equation}
			which can be easily computed as before, leading to:
			\begin{equation}
				\frac{\Delta\mathcal{P}(x_0)}{\mathcal{P}_0}=4\epsilon_0\sqrt{\frac{\pi}{\alpha}}\left(\frac{\xi}{m_\sigma/H}\right)^2 \,\left[-\cos 2 x_0+\left( x_0-\frac{1}{2 x_0}\right)\sin 2 x_0\right]e^{-x_0^2/\alpha},
				\label{eq:Turn_DP}
			\end{equation}
			where now $\alpha=72/\Delta N^2=4\epsilon_0/\Delta^2$. As before, we have that the maximum amplitude is at $x_0=\sqrt{\alpha/2}$. Approximating trigonometric terms in the expression evaluated at $x_0=\sqrt{\alpha/2}$ to 1, the maximum amplitude thus becomes:
			\begin{equation}
				\frac{\Delta \mathcal{P}}{ \mathcal{P}_0} = 4 \epsilon_0\sqrt{\frac{\pi}{2 e}}\left(\frac{\xi}{m_\sigma/H}\right)^2.
				\label{}
			\end{equation}

			As mentioned above, unless $\Delta N\ll1$, the analytical solution matches the numerics quite well, as can be seen by the bottom-center and right panels of Fig.~\ref{fig:turn_analytical}. Compared to the amplitude in the case of the bump/dip, there are two main differences. First, $\alpha_{\rm turn}=2\alpha_{\rm bump/dip}$, so, for the same $\Delta N$, the exponential suppression $\exp(-x_0^2/\alpha)$ is sharper for the bump/dip model. Second, while for the bump/dip model the amplitude switches sign depending on whether we have a bump or a dip in the potential, in the turn model the amplitude is proportional to $\xi^2$, so it always has the same sign.

			\begin{figure}
				\begin{center}
					\includegraphics[width=.327\columnwidth]{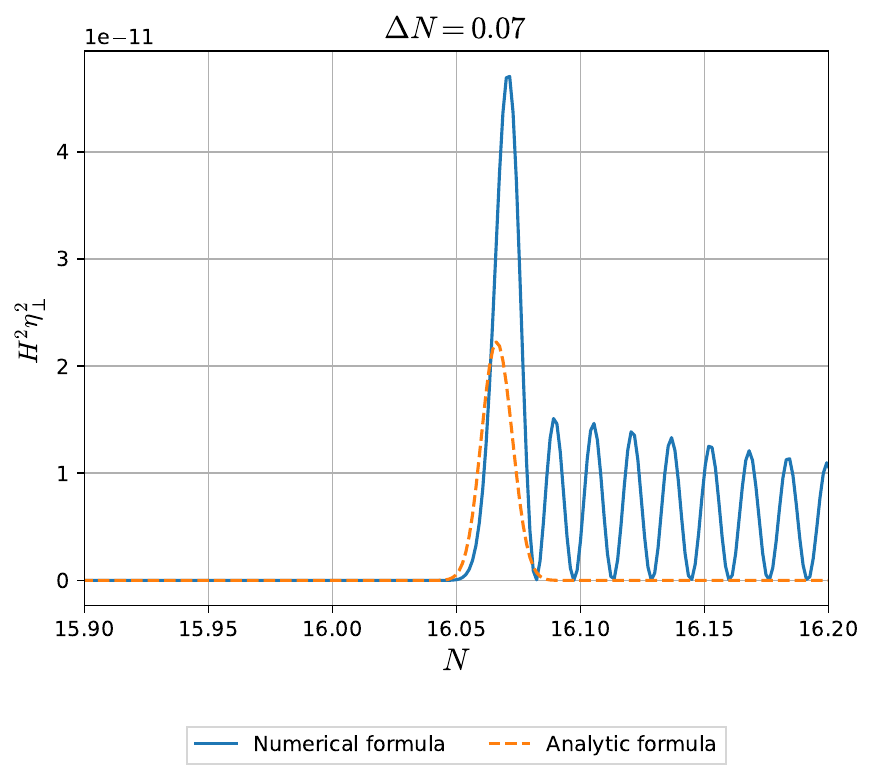}
					\includegraphics[width=.327\columnwidth]{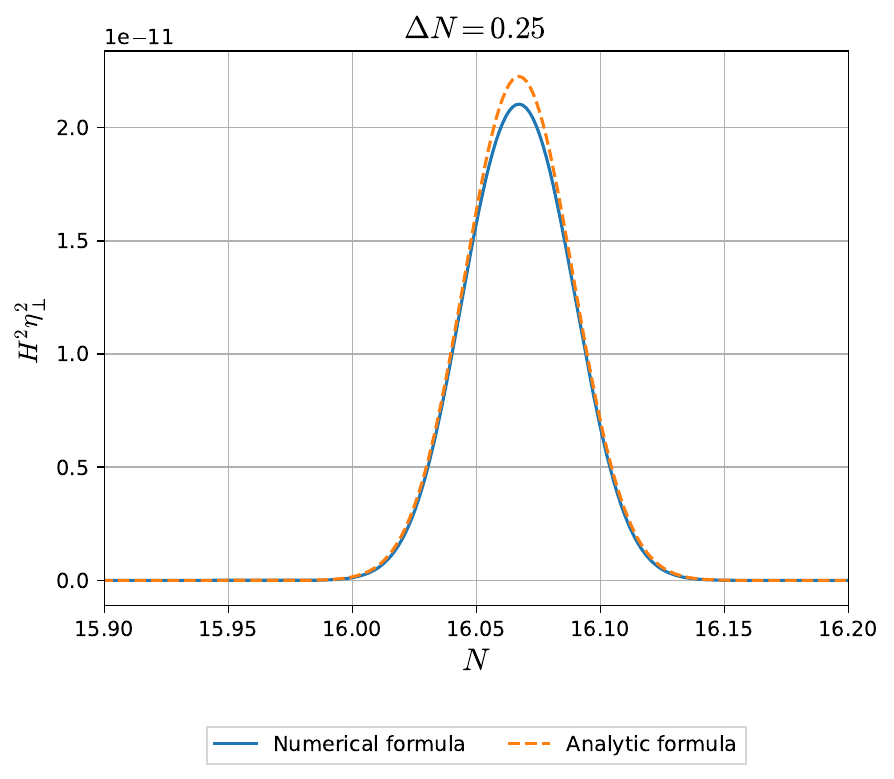}
					\includegraphics[width=.327\columnwidth]{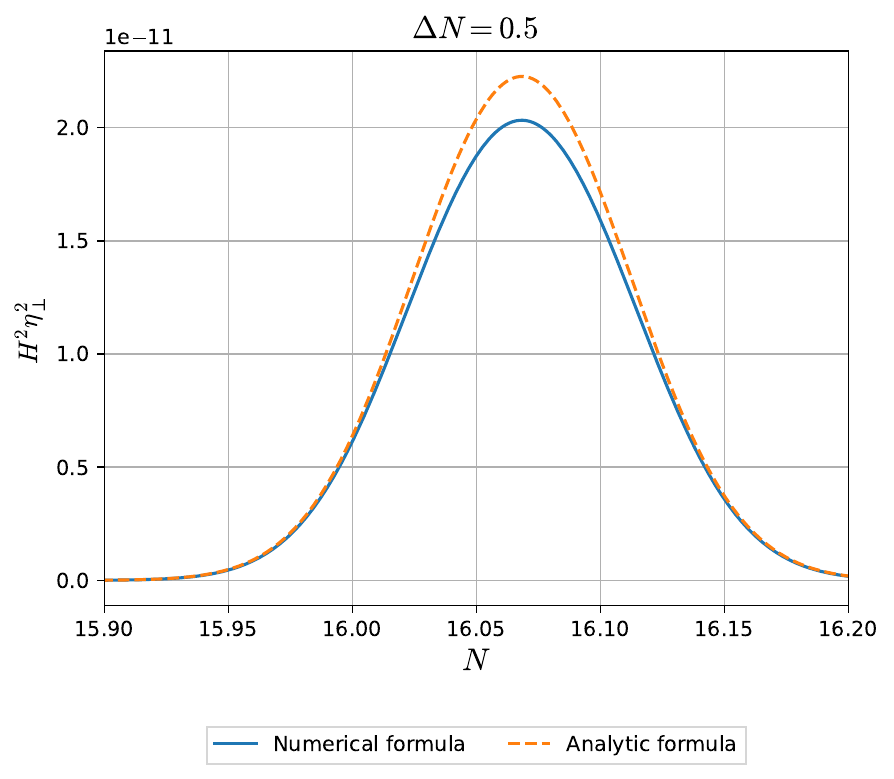}
					\includegraphics[width=.327\columnwidth]{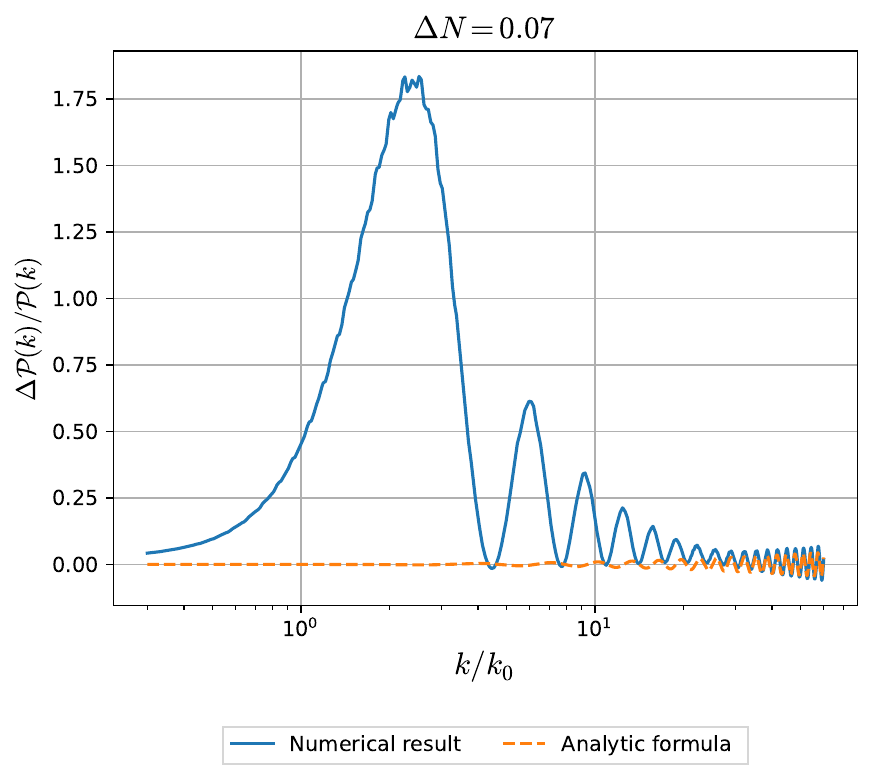}
					\includegraphics[width=.327\columnwidth]{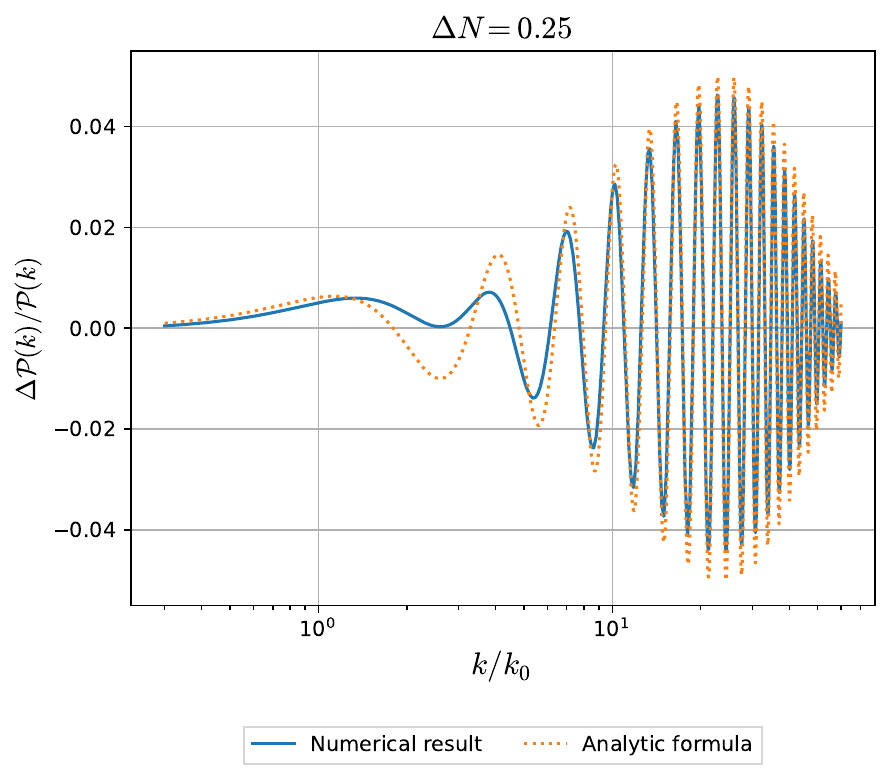}
					\includegraphics[width=.327\columnwidth]{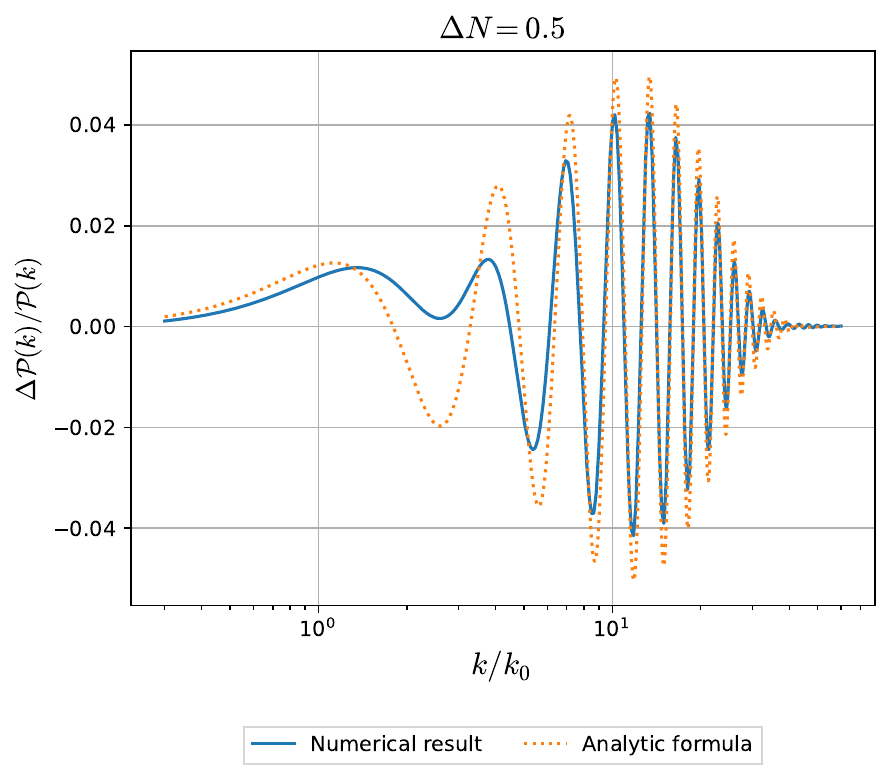}			\end{center}
				\caption{\label{fig:turn_analytical} Comparison of the analytical formulae Eq.~\eqref{eq:etaperp} and \eqref{eq:Turn_DP}. 
				}
			\end{figure}

			\section{Estimate of feature signal in the CMB angular spectra}
			\label{app:cmb_estimates}
			In this Appendix, we present a rough estimate of the effects of primordial features on the CMB angular power spectra. Although the derivation below makes use of several quite crude approximations, the results are very useful to intuitively understand how primordial features are processed by the convolution with the CMB transfer functions and how the $C_\ell$'s are sensitive to different running of primordial features.
			
			The CMB angular power spectra are obtained by convolving the primordial power spectrum with transfer functions describing the late time evolution of the Universe:
			\begin{equation}
				C_{\ell}^{X\, Y}=\int \mathrm{d}\, \log k \, T^{XY}_\ell (k) \mathcal{P}(k)
			\end{equation}
			where $X,\,Y=T,\, E$ and $T^{XY}_\ell (k)$ are the transfer functions, which are integrals along the line of sight of CMB photons of the source terms of $T$ and $E$ modes. The transfer functions are function of $k$ and are labelled by the multipole $\ell$. Some illustrative examples are shown in Fig.~\ref{fig:transfer}.
			The transfer functions are strongly peaked at $\ell\simeq k D_*$, where $D_*$ is the comoving angular diameter distance from the observer to the redshift of recombination. Furthermore, it is important to realize that the E-modes transfer functions are more peaked than the T-modes ones. This is why E-modes are so effective in constraining features, as we will see shortly.

			To estimate the feature signal in the CMB spectra, we will make a series of rough assumptions that are nevertheless reasonable for understanding the main physics. All of them stem from the peaked shape of the transfer functions.
			\begin{itemize}
				\item We will assume the one-to-one relation $\ell(k)= k D_*$ to be always valid. Equivalently $k(\ell)= \ell/ D_*$.
				\item We model the transfer functions as top-hat functions in $\log k$ space as
				\begin{equation*}
					T^{XY}_\ell (k)= T_{\ell,\,{\rm max}}^{XY}\,\Theta\left[\log k - \left(\log k(\ell)\,-\,\Delta_\ell^{XY}/2\right)\right]\,\Theta\left[-\log k + \left(\log k(\ell)\,+\,\Delta_\ell^{XY}/2\right)\right],
				\end{equation*}
with an $\ell$-dependent width $\Delta_\ell^{XY}$, which is also different for each pair $XY$.				This assumption is quite reasonable as, again, the transfer functions are strongly peaked around $\ell(k)$. We show the transfer functions computed with the Einstein-Boltzmann solver for for $X,\,Y=T,\,T$ and their top-hat approximation for $\ell=2$ in the top panel of Fig.~\ref{fig:transfer}. 
			\end{itemize}
			
			Roughly speaking, at a given  $\ell$ (or $k$), if the variation of the  primordial spectrum is slow in the log-interval $\Delta_\ell^{XY}$, its effect will show up in the angular spectrum $C_\ell^{XY}$. On the other hand, if, within  $\Delta_\ell^{XY}$, the PPS oscillates fast, the oscillations will be washed out in the integral over $\ln k$. We now proceed to give a more detailed proof of this naive argument.
			\begin{figure}
				\includegraphics[width=.9\columnwidth]{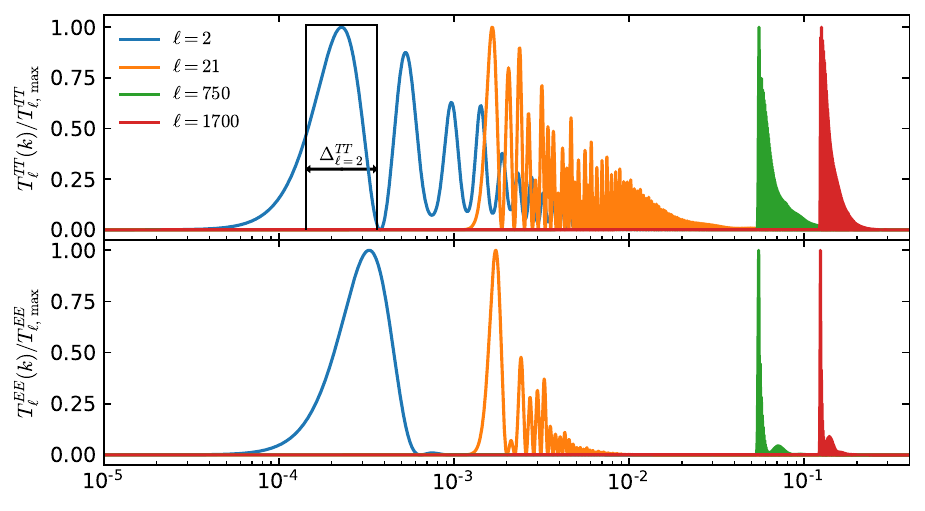}
				\caption{\footnotesize\label{fig:transfer} CMB transfer functions for T-modes (top) and E-modes (bottom). }
				
			\end{figure}

			For simplicity, we focus here on features with a constant envelope. 
			Nevertheless, as clear from the argument above, as long as the ($k$-dependent) envelope is smooth enough over $\Delta_\ell$, the calculation below can be straightforwardly generalized by simply multiplying our results by the envelope. We parameterize the PPS as
			
			\begin{equation}
				\mathcal{P}(k)=\mathcal{P}_0(k)(1+\delta_{\mathcal{P}})
			\end{equation}
			where $\mathcal{P}_0(k)=A_s\left(k/k_*\right)^{n_s-1}$ and

			\begin{equation}
				\label{eq:deltaP}
				\delta_\mathcal{P}(k)=C\begin{cases}\,\,  \sin \omega\log\left(k/k_p\right)& \text{resonant} \\ \,\,\sin k/k_0  & \text{sharp} \end{cases},
			\end{equation}
			For simplicity, we neglect the phases and any $k$-dependent envelops of the features in our calculation.
			
			Our goal is to estimate the correction to the CMB spectra due to $\delta_\mathcal{P}$. To do that, we write 
			\begin{equation}
				C_{\ell}^{X\, Y}\equiv
				C_{\ell}^{X\, Y,\,0}(1+\delta_\ell)=\int \mathrm{d}\, \log k \, T^{XY}_\ell (k) \mathcal{P}_0(k)(1+\delta_{\mathcal{P}})
			\end{equation}

			The calculation of the leading order contribution is very simple. Since $\mathcal{P}_0(k)$ is almost constant, it can be factored out the integral and we are left with:
			
			\begin{align}
				C_{\ell}^{X\, Y,\,0}&=\int \mathrm{d}\, \log k \, T^{XY}_\ell (k) \mathcal{P}_0(k)\notag\\&=T_{\ell,\,{\rm max}}^{XY} \mathcal{P}_0(\ell)\int \mathrm{d} x\, \Theta\left[x - \left(\log k(\ell)\,-\,\Delta_\ell^{XY}/2\right)\right]\,\Theta\left[-x + \left(\log k(\ell)\,+\,\Delta_\ell^{XY}/2\right)\right]\notag\\
				&=T_{\ell,\,{\rm max}}^{XY}\,A_s\left(\frac{\ell}{\ell_*}\right)^{n_s-1}\Delta_\ell^{XY}.
			\end{align} 
			For example, at large scales, we have the very well known Sachs-Wolfe plateau for which $T_{\ell,\,{\rm max}}^{TT}\,\Delta_\ell^{TT}\propto 1/\ell(\ell+1)$. Anyway, the dominant contribution $C^0$ is not relevant for our discussion.
			
			We are interested in $\delta_\ell$. We have
			
			\begin{align}
				C_{\ell}^{X\, Y,\,0}\,\delta_\ell&=T_{\ell,\,{\rm max}}^{XY} \mathcal{P}_0(\ell) \int \mathrm{d}\, \log k \,\cdots\notag\\
				&=T_{\ell,\,{\rm max}}^{XY} \mathcal{P}_0(\ell) \Delta_\ell^{XY} \frac{1}{\Delta_\ell^{XY}}\int \mathrm{d}\, \log k \,\cdots\notag\\
				&= C_{\ell}^{X\, Y,\,0}\frac{1}{\Delta_\ell^{XY}}\int \mathrm{d}\, \log k \,\cdots,
			\end{align} 
			so
			\begin{equation}
				\label{eq:deltal}
				\delta_\ell=\frac{C}{\Delta_\ell^{XY}}\int^{\log k_\ell+\Delta/2}_{\log k_\ell-\Delta/2} \mathrm{d} x\,\,\begin{cases}
					\sin (\omega x) & \text{resonant}  , \\
					
					\sin (e^x / k_0) & \text{sharp} .
				\end{cases} ,
			\end{equation} 
			where, for simplicity, we forgot the $X,\,Y,\,\ell$ labels on $\Delta$ and denoted $k_\ell\equiv k(\ell)$. The effects of sharp and resonant features with constant envelope $C=0.1$ are shown in Fig.~\ref{fig:cmb}. In the following subsections we discuss each of them separately. Let us stress that, in the case where the features have a $k$-dependent envelope, the result above still applies by substituting $C$ with $C(k)$.

			\begin{figure}
				\includegraphics[width=.5\columnwidth]{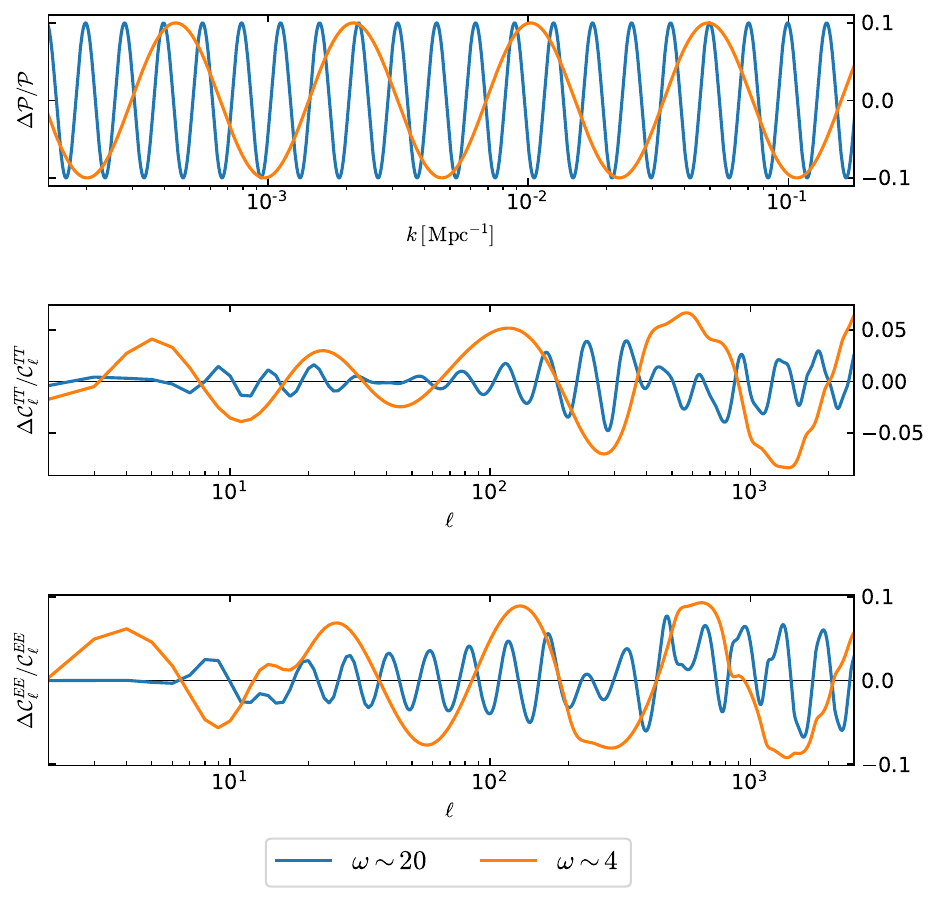}
				\includegraphics[width=.5\columnwidth]{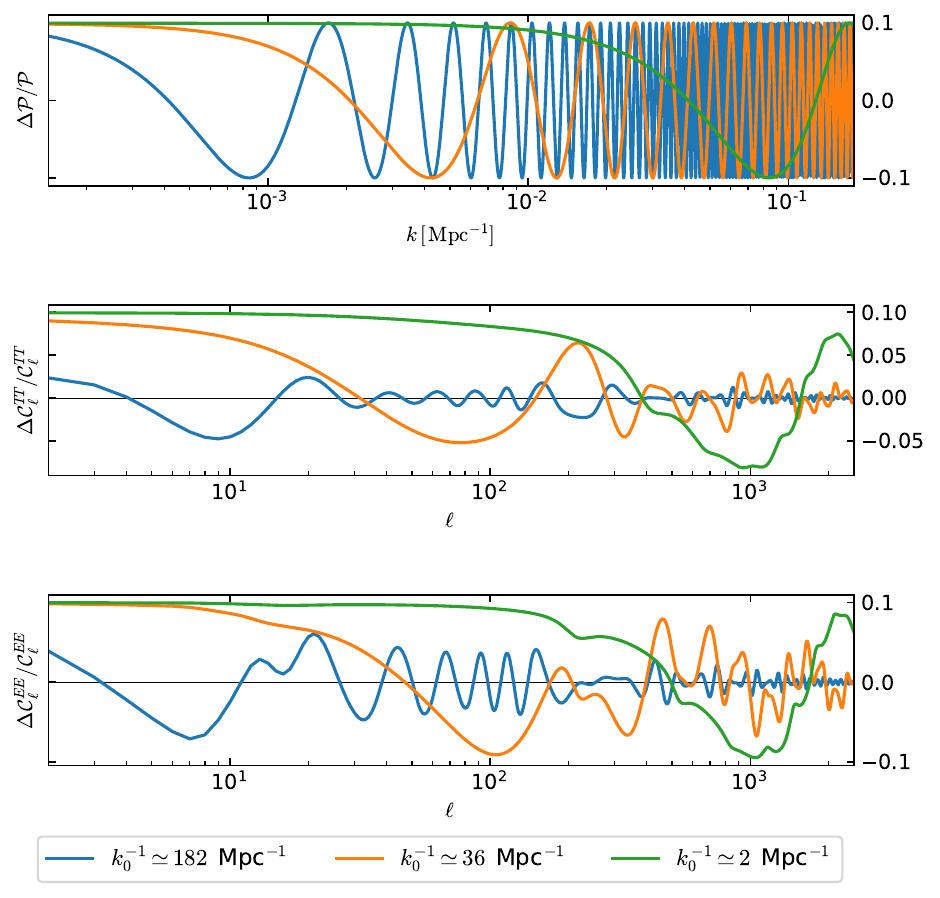}
				\caption{\footnotesize\label{fig:cmb} Residuals due to the $\sin\log\,k$ (left) and $\sin\, k$  (right)  features in the PPS. Note that we keep the cosmological parameters fixed, so the correction is only due the features in the PPS.}
				
			\end{figure}
			
			\subsection{Resonant features: $\sin \log k$}
			Solving the integral~\eqref{eq:deltal}, we obtain:
			
			\begin{equation}
				\label{eq:deltares}
				\delta_\ell^{\rm res}\simeq C \sin \left[\omega \log\,k(\ell)/k_p \right]\,\frac{\sin \left(\omega\Delta_\ell\, /2\right)}{\omega\Delta_\ell \,/2}.
			\end{equation}
			In the limit where the transfer functions are very narrow, i.e. $\omega\Delta_\ell\to0$ at all multipoles, the correction is simply the same as the correction to the PPS $\delta_\ell^{\rm res}\simeq C \sin \left(\omega \log\,k(\ell)/k_p \right)$. This can also be obtained by simply setting the transfer function in the original integral as a Dirac delta. 
			In Fig.~\ref{fig:cmb}, we show an example that satisfies the condition $\omega\,\Delta_\ell/2\ll1$ with $\omega=4$ and $X,\,Y=E,\,E$. See the orange line of the bottom-left panel of Fig.~\ref{fig:cmb}. Note that the plots show the numerical result for $\delta_\ell$ obtained with the Einstein-Boltzmann solver.
			
			If the limit $\omega\,\Delta_\ell/2\ll1$ is not satisfied, as for all the other cases in the left panel of the figure, Eq.~\ref{eq:deltares} can still be used to understand the impact of resonant features on the CMB spectra. Remember that, at a given $\ell$, $\Delta^{TT}_\ell>\Delta^{EE}_\ell$ and $\Delta^{XY}_\ell$ decreases with $\ell$. Therefore, the amplitude of the feature is less washed out in $EE$; and, at a given feature frequency $\omega$, $\delta_\ell$  increases with $\ell$. If we instead fix $\ell$, we see that $\delta_\ell$ is larger for smaller frequencies $\omega$. 
			
			To summarize, the sensitivity to resonant features:
			\begin{itemize}
				\item increases with $\ell$
				\item decreases with $\omega$.
			\end{itemize} 
			As we will see in the next subsection, the first of this conclusions does not apply to sharp features, for which, in fact, the opposite holds.
			
			\subsection{Sharp features: $\sin  k$}
			Solving the integral~\eqref{eq:deltal}, we obtain:
			
			\begin{equation}
				\label{eq:deltasharp}
				\delta_\ell^{\rm sharp}\simeq C \frac{{\rm Si} \left(k\,\frac{e^{\Delta_\ell/2}}{k_0}\right)-{\rm Si} \left(k\,\frac{e^{-\Delta_\ell/2}}{k_0}\right)}{\omega\Delta_\ell \,/2}.
			\end{equation}
			where ${\rm Si}$ is the sine integral function. Also here, in the limit where the transfer functions are very narrow, i.e. $\Delta_\ell\to0$ at all multipoles, the correction is simply the same as the correction to the PPS $\delta_\ell^{\rm res}\simeq C \sin \left(k(\ell)/k_0 \right)$. 
			Some examples are shown in the right panel of Fig.~\ref{fig:cmb}.
			
			In the limit of large frequencies $\omega_{\rm lin}\equiv k_0^{-1}\ll1$ we can expand the formula above as

			\begin{equation}
				\delta_\ell^{\rm sharp}\simeq  \frac{C}{ \Delta_\ell\,k/k_0}\left[e^{\Delta_\ell/2}\cos\left(e^{-\Delta_\ell/2}k/k_0\right) - e^{-\Delta_\ell/2}\cos\left(e^{\Delta_\ell/2}k/k_0\right)\right].
			\end{equation}
			The envelope of $\delta_\ell$ is very well approximated by $\sim2 e^{\Delta_\ell/2}k_0/k \Delta_\ell =2e^{\Delta_\ell/2}/\Delta_\ell \omega_{\rm lin}\,k$. The consequence is that $\delta_\ell$ decreases at smaller scales because of the suppression factor $k^{-1}$. Note that $\delta_\ell$ increases too because $\Delta_\ell^{-1}$ grows for large wavenumbers $k$, but it does not grow enough to compensate for the $k^{-1}$ factor. Furthermore $\delta_\ell$ is suppressed for large values of $\omega_{\rm lin}$, as in the case for resonant features (see above). 
			
			To summarize, the sensitivity to sharp features:
			\begin{itemize}
				\item decreases with $\ell$
				\item decreases with $\omega$.
			\end{itemize}

\bibliographystyle{JHEP}
\bibliography{main}

\end{document}